\documentclass[compsoc,conference,a4paper,10pt,times]{IEEEtran}
\IEEEoverridecommandlockouts
\usepackage{cite}
\usepackage{amsmath,amssymb,amsfonts}
\usepackage{algorithmic}
\usepackage{graphicx}
\usepackage{textcomp}
\usepackage{bmpsize}
\usepackage{xcolor}
\usepackage{lipsum}

\usepackage{amsmath}
\usepackage{amssymb,amsfonts}
\usepackage{algorithmic}
\usepackage{graphicx}
\usepackage{textcomp}
\usepackage{xcolor}

\usepackage{multirow}
\usepackage{bm}
\usepackage{stfloats}
\usepackage{makecell}
\usepackage{soul}
\usepackage{float}

\usepackage{enumitem}
\usepackage{tcolorbox}

\usepackage{pifont}
\usepackage{fontawesome5}
\usepackage{mathrsfs}

\usepackage[colorlinks=true,urlcolor=black]{hyperref}
\def\BibTeX{{\rm B\kern-.05em{\sc i\kern-.025em b}\kern-.08em
    T\kern-.1667em\lower.7ex\hbox{E}\kern-.125emX}}
\begin{document}

\title{SecDTD: \underline{D}ynamic \underline{T}oken \underline{D}rop for \underline{Sec}ure Transformers Inference}

\author{
\IEEEauthorblockN{
Yifei Cai$^{\dagger}$, 
Zhuoran Li$^{*}$, 
Yizhou Feng$^{\ddagger}$, 
Qiao Zhang$^{\S}$, 
Hongyi Wu$^{*}$, 
Danella Zhao$^{*}$, 
and Chunsheng Xin$^{\dagger}$
}
\IEEEauthorblockA{
$^{\dagger}$Iowa State University, yifeic@iastate.edu, cxin@iastate.edu \\
$^{*}$University of Arizona, zli1122@arizona.edu, mhwu@arizona.edu, danellazhao@arizona.edu \\
$^{\ddagger}$Old Dominion University, yfeng002@odu.edu \\
$^{\S}$Shandong University, qiao.zhang@sdu.edu.cn
}

\thanks{This work has been accepted for publication at the 11th IEEE European Symposium on Security and Privacy (EuroS\&P 2026).}

}

\maketitle

\begin{abstract}
The rapid adoption of Transformer-based AI has been driven by accessible models such as ChatGPT, which provide API-based services for developers and businesses. 
However, as these online inference services increasingly handle sensitive inputs, privacy concerns have emerged as a significant challenge. To address this, secure inference frameworks have been proposed, but their high computational and communication overhead often limit practical deployment. 
In plaintext settings, token drop is an effective technique for reducing inference cost; however, our analysis reveals that directly applying such methods to ciphertext scenarios is suboptimal due to distinct cost distributions in secure computation.
We propose SecDTD, a dynamic token drop scheme tailored for secure Transformer inference. SecDTD advances token drop by shifting the dropping to earlier inference stages, effectively reducing the cost of key components such as Softmax. To support this, we introduce two core techniques.
Max-Centric Normalization (MCN): A novel, Softmax-independent scoring method that enables early token drop with minimal overhead and improved normalization, supporting more aggressive dropping without accuracy loss.
OMSel: A faster, oblivious median selection protocol that securely identifies the median of importance scores to support token drop. Compared to existing sorting-based methods, OMSel achieves a 16.9$\times$ speedup while maintaining security, obliviousness and randomness.
We evaluate SecDTD through 48 experiments across eight GLUE datasets under various network settings using the BOLT and BumbleBee frameworks. SecDTD achieves 4.47$\times$ end-to-end inference acceleration without degradation in accuracy. 
\end{abstract}


\section{Introduction}
\label{introduction}

Transformer-based models, such as BERT~\cite{devlin2019bert} and GPT~\cite{radford2018improving,radford2019language}, have revolutionized artificial intelligence (AI) by achieving state-of-the-art performance in tasks like language translation \cite{ott2018scaling}, question answering \cite{karpukhin2020dense}, and code generation \cite{chen2021evaluating}. Their success is largely attributed to attention mechanisms \cite{vaswani2017attention}, which effectively capture contextual relationships and long-range dependencies. Unlike traditional task-specific supervised learning, large Transformers undergo pre-training on massive amounts of unlabeled data, enhancing their adaptability across diverse applications.

The widespread adoption of Transformer-based AI has been further driven by publicly accessible models like ChatGPT \cite{ChatGPT}, which offer API services \cite{10.14778/3282495.3282499} for developers and businesses. This accessibility has facilitated their integration into critical domains such as healthcare \cite{act1996health,HIPAAcompliance2025}, where Transformers assist in analyzing medical records and extracting insights to support diagnosis. However, as these models increasingly process sensitive data, privacy concerns have become more pressing \cite{denecke2024transformer}. Online inference services pose risks, as user-submitted prompts may contain personal information, raising ethical and security challenges. Addressing privacy concerns is essential for the responsible deployment of Transformer models in sensitive applications.

To address privacy risks in AI inference, various privacy preserving frameworks have been developed to protect model parameters from users while preventing servers from accessing users’ private inputs. Recent research has primarily focused on privacy-preserving frameworks based on Secure Multi-Party Computation (MPC) \cite{goldreich2019play,10.1145/3335741.3335756} and Homomorphic Encryption (HE)~\cite{paillier1999public,brakerski2014efficient} to enable private inference on deep learning models. These frameworks are applied to convolutional neural networks (CNNs) \cite{liu2017oblivious,gilad2016cryptonets,rathee2020cryptflow2,244032,juvekar2018gazelle,279898} and Transformer-based models \cite{hao2022iron,pang2023bolt,lu2023bumblebee,zhang2024secure}, ensuring secure computation while preserving model parameters in privacy-preserving inference.

\vspace*{0.05in}
\noindent\textbf{Efficiency Challenges for Privacy-preserving Transformers Inference: }
While secure frameworks enable encrypted computation, they often incur substantial computation and communication costs due to the reliance on expensive cryptographic primitives \cite{cai2022hunter,cai2024mosaic,zhang2024individual}. Privacy-preserving Transformer frameworks face two key challenges:
First, unlike CNN inference, which primarily involves convolution operations and relatively simple matrix-vector multiplications, Transformers heavily rely on large-scale matrix-matrix multiplications \cite{vaswani2017attention}. These operations dominate the \textbf{linear} computational cost and significantly increase the resource demands of secure inference.
Furthermore, in contrast to CNNs that typically use the lightweight ReLU activation function, Transformers require the secure evaluation of numerous complex \textbf{non-linear} functions such as Gaussian Error Linear Unit (GeLU), Softmax, and LayerNorm. These functions must be computed over hundreds of thousands of inputs, substantially increasing both computational and communication overhead in privacy-preserving inference \cite{rathee2021sirnn,rathee2020cryptflow2}.

Recent privacy-preserving Transformer frameworks have made notable progress in improving inference efficiency. However, secure Transformer inference remains highly resource-intensive. For instance, executing BERT-base on a single input from the MRPC dataset can exceed 1,000 seconds under WAN network conditions \cite{hao2022iron,pang2023bolt}, rendering it impractical for real-world deployment. Addressing this inefficiency is therefore critical.

\vspace*{0.05in}
\noindent\textbf{Existing Token Drop Schemes and Limitation:}
In plaintext scenarios, a common strategy for reducing Transformer inference costs involves eliminating redundant tokens from the input sequence. For instance, PowerBERT \cite{goyal2020power} introduces a method that assesses the significance of word vectors based on Softmax outputs, guiding the dropping process to retain only the most relevant tokens. This approach effectively improves the efficiency of both linear and non-linear operations during plaintext inference.
To extend this strategy to secure inference, \cite{pang2023bolt} proposed a straightforward Word Elimination (W.E.) method. Subsequently, CipherPrune \cite{zhang2025cipherprune} introduced a token dropping technique that adaptively removes low-importance tokens based on a pre-learned significance threshold.

\begin{figure}[t]
\centering
\includegraphics[trim={0cm 0cm 0cm 0cm}, clip, scale=0.38]{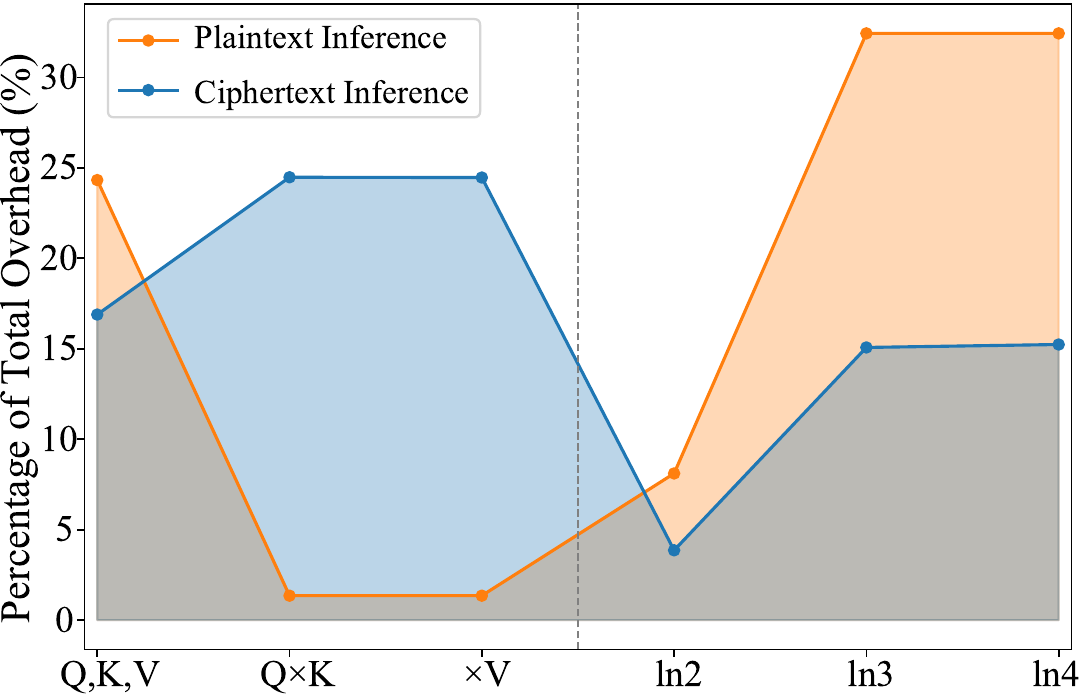}
\vspace*{-0.1in}
\caption{Cost Distribution Comparison Between Plaintext and Ciphertext Transformer Inference (BERT-base, Input Sequence Length of 128).} 
\vspace*{-0.15in}
\label{pt_vs_ct}
\end{figure}

However, as our analysis indicates, existing secure token drop strategies are still suboptimal:

$\bullet$ \textbf{Inadequate token drop coverage in ciphertext scenarios.}
Existing secure token drop strategies are largely inherited from their plaintext counterparts. This key limitation arises from fundamental differences in cost distribution between plaintext and ciphertext Transformer inference.
Referring to Figure~\ref{Overview}, the Transformer typically proceeds through several key components to complete the inference process. The linear operations occur sequentially as Q,K,V; Q$\times$K; $\times$V; ln2; ln3; and ln4. (The detailed explanation of each component can be found in Section~\ref{Secure_Transformer}.)
As illustrated in Figure~\ref{pt_vs_ct}, the cost distribution differs significantly between the two settings. In the plaintext scenario (shown in orange), the inference cost is primarily concentrated in ln2, ln3, and ln4, which together account for \textbf{73\%} of the total cost. This observation drives the design of plaintext-based token drop method, where token removal is strategically applied after the $\times$V operation and before ln2 to maximize cost savings.
In contrast, in the ciphertext scenario (shown in blue), the inference cost shifts toward earlier stages, with ln2, ln3, and ln4 contributing only \textbf{34\%} of the total cost. As a result, token drop strategies based on plaintext cost patterns become substantially less effective when directly applied to ciphertext settings.
Additionally, nonlinear operations such as Softmax—which are computationally lightweight during plaintext inference—incur significant overhead in ciphertext scenarios, further exacerbating the inefficiency of existing approaches. This inefficiency is compounded by the fact that both plaintext token drop methods and existing ciphertext token drop methods rely on Softmax outputs to assess token significance. As a result, Softmax itself cannot benefit from token dropping.

$\bullet$ \textbf{High-overhead and low-efficiency token drop decision strategy.}
Another critical aspect is deciding which tokens to drop. Work W.E. employs Bitonic Sort to find the median score of input sequence and discards tokens with scores below the median.
However, this approach incurs high complexity in MPC environments; for example, fully sorting a 256-element vector requires 13,824 operations, making it costly for real-time inference. 
CipherPrune uses a static threshold learned through fine-tuning, dropping tokens with scores lower than this fixed threshold. This threshold-based approach is commonly used in model pruning \cite{han2015learning}.
While the comparison with a fixed threshold is relatively inexpensive, learning an appropriate threshold requires costly fine-tuning. Moreover, the learned threshold tends to be dataset-dependent and lacks adaptability to diverse real-world inputs. As a result, when switching tasks or datasets, additional fine-tuning is typically necessary.
In addition, static thresholding is not efficient when using packed HE primitives, which are commonly adopted in privacy-preserving AI. This is because different inputs may retain varying numbers of tokens after applying a static threshold. While such variability is generally non-critical in plaintext scenarios, it becomes problematic in ciphertext settings, where the inability to precisely control the number of retained tokens leads to suboptimal ciphertext packing.
For example, if 16 tokens fit exactly into a single ciphertext, reducing a sequence from 32 to 16 tokens decreases the number of ciphertexts from two to one. However, if 17 tokens remain after static thresholding, two ciphertexts are still needed. This underscores the importance of precise control over the number of retained tokens to maximize packing efficiency.

$\bullet$ \textbf{Potential leakage concerns.}
During token dropping, the token contents are encrypted, and the specific tokens being dropped are hidden from both parties as well. However, some token drop methods may still pose a risk of information leakage.
For example, when use static thresholding, different inputs typically lead to different dropping decisions, that is, varying numbers of tokens are retained. Moreover, the same input consistently results in the same dropping pattern. This creates a one-to-one mapping between each input and its corresponding dropping decision, forming a unique ``fingerprint" for each input. Although the content of the input remains encrypted, these ``fingerprints" may leak side-channel information and thus introduce potential privacy risks.

\begin{figure}[t]
\centering
\includegraphics[trim={0cm 0cm 0cm 0cm}, clip, scale=0.54]{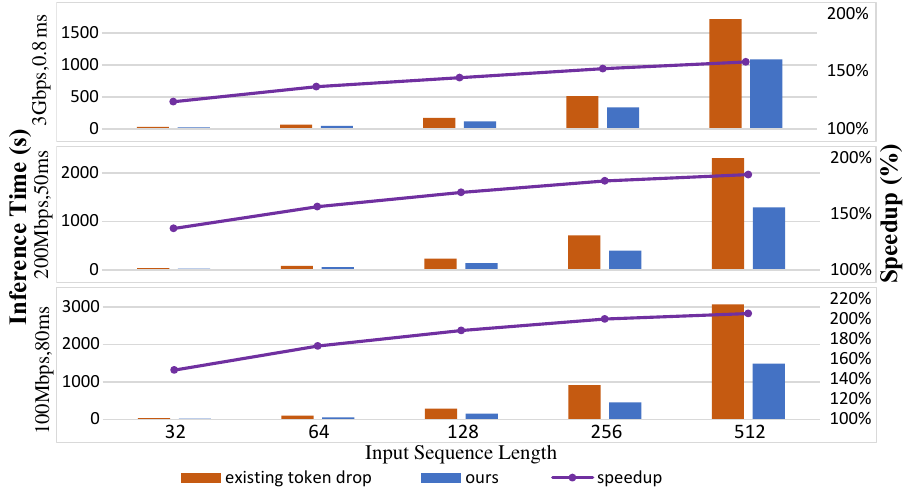}
\vspace*{-0.25in}
\caption{Comparison of Inference Time between Existing Token Drop Method and our SecDTD with “Pre-Softmax Gain”.} 
\vspace*{-0.1in}
\label{vs_W.E}
\end{figure}

These challenges motivate us to develop a Dynamic Token Drop methodology for encrypted Transformer Inference, \textbf{SecDTD}, which is more efficient, more secure, and incurs lower operational cost.
SecDTD introduces the following \textbf{key contributions}:

\vspace*{0.05in}\noindent\textbf{Pre-Softmax Gain Token Drop.}
SecDTD strategically shifts the token drop operation to earlier stages of Transformer inference, aligning with the unique cost distribution in ciphertext-based scenarios. SecDTD not only benefits later layers such as ln2, ln3, and ln4, but also extends the advantages of token drop to earlier operations, including Softmax, $\times V$, and even the initial computation of $I \times W_v = V$. This ``pre-Softmax gain” effectively reduces the runtime cost of multi-head attention.
As shown in Figure~\ref{vs_W.E}, we compare SecDTD with existing token drop method under three network conditions, varying the input sequence length from 32 to 512 tokens. Assuming that half of the tokens are dropped, SecDTD consistently outperforms the baseline across all configurations, due to the ``pre-Softmax gain''. In scenarios involving longer input sequences and more constrained network conditions, which closely resemble practical deployments, our approach demonstrates substantial advantages, achieving up to a \textbf{206\%} speedup.
Even greater gains can be expected as input lengths continue to grow in real-world applications.

\vspace*{0.05in}\noindent\textbf{Softmax-independent Scoring Method, MCN.}
To evaluate the importance of each token in the input sequence, existing scoring methods typically rely on the output of Softmax. This dependency prevents earlier stages, including Softmax itself, from benefiting from token dropping.
To achieve the pre-Softmax gain token drop, we introduce a novel scoring method, Max-centric Normalization (MCN), which assesses token importance with three key advantages:
\textbf{(1)} It is independent of Softmax, enabling token drop before Softmax, thus benefiting earlier stages.
\textbf{(2)} The MCN algorithm incurs negligible additional overhead in MPC environments.
\textbf{(3)} Unlike W.E.'s scoring mechanism, which lacks normalization, MCN incorporates normalization to better capture token importance while mitigating the impact of extreme values. This allows multiple rounds of token drop and supports high-ratio token drop without significant accuracy degradation, a challenge faced by W.E.

\vspace*{0.05in}\noindent\textbf{Fast Oblivious Median Selection, OMSel.}
Once token importance scores are obtained, determining which tokens to drop becomes crucial. 
\textcolor{black}{Similar to W.E., we adopt a half-and-half token drop strategy, removing half of the current tokens at each interaction by discarding those with scores below the median. However, our approach differs significantly in how the median is computed.}
W.E. relies on Bitonic Sort, which has high complexity in MPC environments due to the complete sorting of token scores. To enhance efficiency, we propose a novel and secure median search method, Oblivious Median Selection (OMSel). 
Compared to W.E.'s sorting-based approach, OMSel achieves up to a \textbf{16.9$\times$} speedup, making token drop practical for complex tasks with long input sequences and multiple token drop rounds.
OMSel efficiently identifies the median while ensuring security and obliviousness. During the median search, neither the Client nor the Server gains information about the actual token scores, the relative rankings of tokens, or intermediate decisions. To maintain obliviousness, all tokens remain engaged in the search process, even if they are predetermined for removal or retention. Additionally, randomness is introduced to obscure access patterns and prevent response-based attacks. 
Importantly, since OMSel consistently drops the lower half of tokens, those below the median, across different inputs, it avoids creating a one-to-one mapping between each input and its dropping decision. This design effectively mitigates the risk of potential information leakage. (Refer to the detailed security analysis in Subsection~\ref{OMSel}.)

\begin{figure}[t]
\centering
\includegraphics[trim={0cm 0cm 0cm 0cm}, clip, scale=0.45]{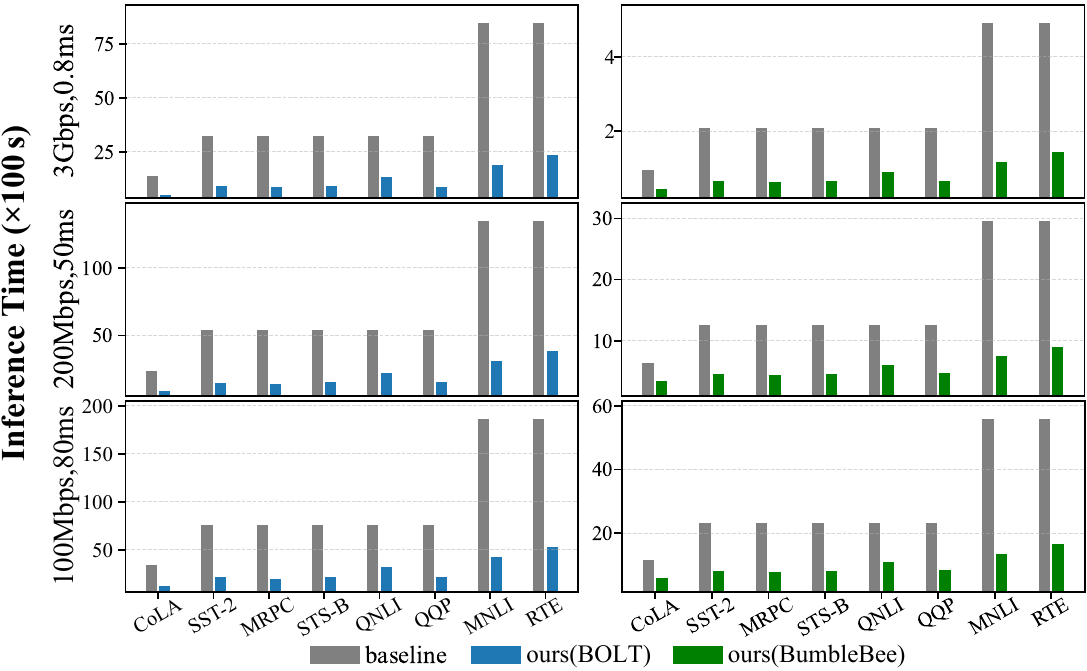}
\vspace*{-0.25in}
\caption{Performance Evaluation of SecDTD across Multiple Datasets, Network Environments and Frameworks.} 
\vspace*{-0.1in}
\label{evals}
\end{figure}

\vspace*{0.05in}
We conduct extensive experiments to validate the efficiency of SecDTD. Evaluations are performed through 48 sets of experiments on BERT-base across eight datasets: CoLA, SST-2, MRPC, STS-B, QNLI, QQP, MNLI, and RTE, covering various input lengths and task types. These evaluations span three network environments and are implemented within both the BOLT and BumbleBee frameworks\footnote{In this paper, the two frameworks employ different levels of thread optimization; please focus on the \textbf{ratio} rather than the absolute values. See Section~\ref{Experimental_Setup} for details.}, as illustrated in Figure~\ref{evals}. SecDTD achieves up to \textbf{4.47$\times$} end-to-end acceleration compared to the baseline, while maintaining model accuracy without any fine-tuning.
With quick fine-tuning (within 5 epochs), further speedups are observed. For example, with fine-tuning, SecDTD accelerates BERT-base inference on RTE from \textbf{3.51$\times$} to \textbf{4.04$\times$} since fine-tuning helps recover accuracy losses caused by more aggressive token drop. 
More details are provided in Section~\ref{Evaluation}.

\section{Preliminaries}
\label{preliminary}

\subsection{Secure Inference and Threat Model}
\label{Secure_Inference}

\begin{figure*}[t]
\centering
\includegraphics[trim={0cm 0cm 0cm 0cm}, clip, scale=0.8]{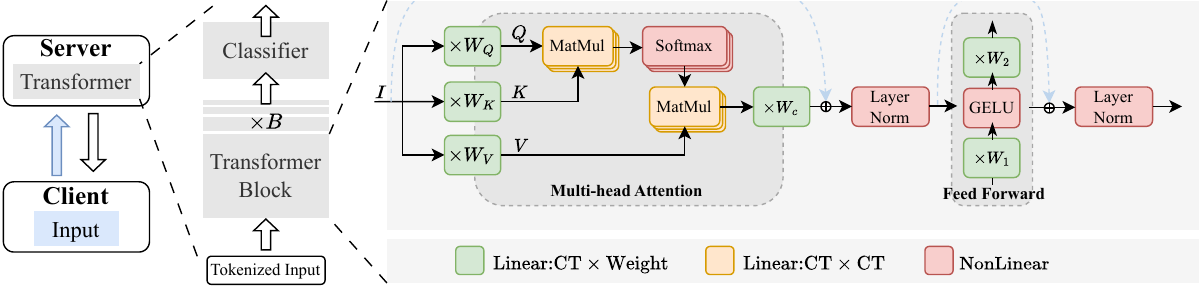}
\vspace*{-0.1in}
\caption{Overview of Secure Transformer Inference.} 
\vspace*{-0.1in}
\label{Overview}
\end{figure*}

Follow the secure inference protocol for Transformer models, such as Iron\cite{hao2022iron}, BOLT\cite{pang2023bolt}, BumbleBee\cite{lu2023bumblebee}, and Nexus\cite{zhang2024secure}. As shown in Figure~\ref{Overview}, the service provider maintains a fine-tuned Transformer model with private weights, offering it privately to users who hold private input data. This represents a typical secure two-party computation scenario. The secure Transformer inference framework enables the client to query the server’s inference service and obtain the model’s output on its input.

Following existing work \cite{juvekar2018gazelle,279898,244032,rathee2020cryptflow2,pang2023bolt,lu2023bumblebee}, we consider privacy protection against a semi-honest adversary, where the two mutually distrusting parties, the server and the client, are honest-but-curious. That is, we assume both parties follow the protocol specification but may attempt to passively infer additional information. Semi-honest security is a common assumption in privacy-preserving machine learning (PPML).
Specifically, we assume that the model architecture is known to both parties. As is standard in secure computation protocols, we do not aim to conceal the inference result, as its disclosure is inherent to the ideal functionality of such protocols—the result must be accessible to at least one party, and often both, to be useful. However, orthogonal techniques such as differential privacy \cite{dwork2014algorithmic} can be integrated with the secure Transformer inference framework to mitigate information leakage.

Notably, since this work serves as a solution for secure dynamic token drop, we proactively incorporate necessary security measures to safeguard the additional functionalities associated with token drop. For example, when analyzing the importance scores of input tokens to compute the median, we introduce a random pivot to enhance randomness and defend against response attacks in which a client may craft specific inputs to infer certain model characteristics. Additionally, we incorporate obliviousness into the median-searching process to obscure the relative importance relationships among tokens. For instance, information such as token-A being more important than token-B remains confidential in our design for both parties. For a detailed security analysis, refer to Subsection \ref{sec_MCN} and Subsection \ref{OMSel}.

\vspace*{-0.05in}
\subsection{Transformer}
\label{Transformer}

\vspace*{-0.05in}
In this work, we focus on BERT~\cite{devlin2019bert}, a widely adopted Transformer-based model. The overall architecture and workflow of BERT are illustrated in Figure~\ref{Overview}. Throughout this paper, we consider the BERT-base configuration, which consists of $B = 12$ Transformer blocks (or layers). Each Transformer layer includes a multi-head attention mechanism followed by a feed-forward neural network.

\textbf{Multi-head Attention:}
The attention mechanism is designed to capture context and dependencies among tokens in the input sequence. Specifically, the input $I \in \mathbb{R}^{m \times d}$ is linearly projected into the Query, Key, and Value, denoted as $Q$, $K$, and $V$, using the parameter matrices $W_Q$, $W_K$, and $W_V$, respectively:
\begin{align}
\label{Q,K,V}
Q &=  I \times W_Q, \nonumber\\
K &=  I \times W_K,\\
V &=  I \times W_V, \nonumber
\end{align}
\vspace*{-0.2in}

The attention output for each head, denoted as $Att_h$, is computed as follows, concatenated across all heads, and then passed to a feed-forward network:
\vspace*{-0.1in}
\begin{align}
Att_h &= Softmax(A_h)\times V_h, 
\label{xV}
\end{align}
where $A_h$ is defined as:
\begin{align}
A_h &= \frac{Q_h \times K_h^T}{\sqrt{d/H}}, h \in [H] 
\label{QxK}
\end{align}
and the Softmax function is applied as follows:
\begin{align}
Softmax(x)_{ij} &=  \frac{e^{x_{ij}}}{\sum_{j \in [m]} e^{x_{ij}}}, i\in [m], j\in [m]
\end{align}
Here, $m$ denotes the sequence length (i.e., the number of tokens). In BERT-base, the model dimension is $d = 768$, and the number of attention heads is $H = 12$.

Furthermore, the attention output is normalized before being fed into the feed-forward. The layer normalization is defined as:
\begin{align}
LayerNorm(x)_{ij} &=  \frac{\gamma_j(x_{ij} - \mu_i)}{\sigma_i} + \beta_j, i\in [m], j\in [d]
\label{LayerNorm}
\end{align}
where $\mu_i$ and $\sigma_i$ denote the mean and standard deviation of the $i$-th token's representation, and $\gamma_j$, $\beta_j$ are learnable parameters.

\textbf{Feed-forward:}
A feed-forward network (FFN) typically consists of two linear transformations with a GELU activation function applied between them, chosen for its favorable curvature and non-monotonicity properties~\cite{hendrycks2016gaussian}:
\begin{align}
FFN(X) &= GELU(XW_1)W_2, 
\label{FFN}
\end{align}
\vspace*{-0.2in}
\begin{align}
GELU(x) &= \frac{1}{2}x \cdot (1+erf(\frac{x}{\sqrt{2}})), 
\label{GELU}
\end{align}
where the Gaussian error function (erf) is defined as:
\begin{align}
erf(x) = \frac{2}{\sqrt{\pi}} \int_0^x e^{-t^2} dt,
\end{align}

\subsection{Cryptographic Primitives}
\label{Cryptographic_Primitives}

\noindent\textbf{Secure Multi-party Computation (MPC).} 
MPC techniques enable a group of parties to jointly evaluate a function over their private inputs without revealing any information about those inputs to each other \cite{goldreich2019play,10.1145/3335741.3335756}.

\vspace*{0.05in}\noindent\textbf{Additive Secret Sharing (SS).}
The MPC framework employed in this work uses 2-out-of-2 additive SS \cite{shamir1979share} between two parties, $P_0$ and $P_1$. Each party $P_i$ ($i \in \{0,1\}$) holds a share $\langle x\rangle^i$ of a secret value $x \in \mathbb{Z}_{2^l}$ such that $x = \langle x\rangle^0 + \langle x\rangle^1 \mod 2^\mathscr{l}$, where $\mathscr{l}$ denotes the bit length. With additive SS, additions can be performed locally without communication, while multiplications require interaction between the parties and are realized using oblivious transfer.

\vspace*{0.05in}\noindent\textbf{Oblivious Transfer (OT).}
OT is used to enable secure evaluation of nonlinear operations \cite{brassard1986all}. In a general 1-out-of-2 OT, denoted as $\left(\begin{smallmatrix} 2\\1\end{smallmatrix}\right)$-OT$_\mathscr{l}$, the sender provides two $\mathscr{l}$-bit messages $m_0$ and $m_1$, and the receiver inputs a choice bit $c \in \{0, 1\}$. At the conclusion of the protocol, the receiver obtains $m_c$, while the sender learns nothing about the receiver's choice \cite{asharov2013more,demmler2015aby,sureshaby2,mohassel2018aby3}.

\vspace*{0.05in}\noindent\textbf{Homomorphic Encryption (HE).}
HE is a foundational cryptographic primitive that enables secure computation by allowing linear operations to be performed directly on ciphertexts without requiring decryption~\cite{albrecht2021homomorphic,brakerski2012fully,brakerski2013packed,paillier1999public,brakerski2014efficient,gentry2009fully}. This property permits computations on encrypted data, ensuring data confidentiality throughout the process. Standard HE schemes support operations such as encryption, decryption, homomorphic addition, and homomorphic multiplication. Some secure frameworks may also necessitate homomorphic rotation operations.

\vspace*{0.05in}
\noindent\textbf{Secure Transformer Inference:}
\label{Secure_Transformer}

\vspace*{0.03in}
\noindent Within private frameworks such as BOLT and BumbleBee, Transformer inference can be divided into several \textbf{Stages ($\mathcal{S}$\_)} and securely evaluated using different cryptographic primitives:

\vspace*{0.03in}
\noindent\textbf{$\mathcal{S}$(Q,K,V):}
Referring to Formula~\ref{Q,K,V}, this operation uses HE multiplication in the form of $\text{CT} \times \text{Weight}$, where $I$ represents the HE-encrypted ciphertext, and $W_Q$, $W_K$, and $W_V$ are weight matrices.

\vspace*{0.03in}
\noindent\textbf{$\mathcal{S}$(Q$\times$K):}
This step is simplified as follows:
\vspace*{-0.03in}
\begin{align}
A &=  Q \times K^T, 
\end{align}
\vspace*{-0.18in}

\noindent In the BOLT framework, this operation uses HE multiplication as $\text{CT} \times \text{CT}$, where both $Q$ and $K$ are HE-encrypted ciphertexts. In the BumbleBee framework, following MPC principles, the multiplication is represented as the product of shared values:
\begin{align}
A &=  Q \times K^T \nonumber\\ 
  &= (\langle Q\rangle^0 + \langle Q\rangle^1)\times(\langle K^T\rangle^0+\langle K^T\rangle^1),\\ \nonumber
  &=  \langle Q\rangle^0  \langle K^T\rangle^0 + \langle Q\rangle^0  \langle K^T\rangle^1 + \langle Q\rangle^1  \langle K^T\rangle^0 \\ \nonumber
  &+ \langle Q\rangle^1  \langle K^T\rangle^1
\end{align}
Here, multiplication is performed using HE in the form of $\text{CT} \times \text{Weight}$, as each party treats its own share as plaintext.
\textit{We adopt the BOLT configuration as the primary presentation in this paper, as illustrated in Figure~\ref{Overview}}. Nevertheless, our proposed scheme remains compatible with the BumbleBee framework, with corresponding evaluations conducted under this setting.

\vspace*{0.03in}
\noindent\textbf{$\mathcal{S}$(Softmax):}
In secure inference, the Softmax function is typically divided into two phases: exponential computation and division by the sum:
\vspace*{-0.1in}
\begin{align}
Softmax(x)_{ij} &=  \underbrace{({e^{x_{ij}}})}_{\textbf{\text{Ph-1}}} \cdot \underbrace{(\frac{1}{\sum_{j \in [m]} e^{x_{ij}}})}_{\textbf{\text{Ph-2}}}, 
\label{softmax_2ph}
\end{align}
Each phase can be implemented using linear approximation methods, as in BOLT \cite{pang2023bolt} and BumbleBee\cite{lu2023bumblebee}, or with Look-Up Table (LUT)-based techniques, as proposed in Iron \cite{hao2022iron} and SIRNN \cite{rathee2021sirnn}.

\vspace*{0.03in}
\noindent\textbf{$\mathcal{S}$($\times$V):}
Following Formula~\ref{xV}, this operation is performed similarly to $\mathcal{S}$(Q$\times$K) using $\text{CT} \times \text{CT}$ multiplication in the BOLT framework and $\text{CT} \times \text{Weight}$ multiplication in the BumbleBee framework.

\vspace*{0.03in}
\noindent\textbf{$\mathcal{S}$(ln2):\label{S:ln2}}
This step concatenates the attention outputs from all heads using HE multiplication in the form of $\text{CT} \times \text{Weight}$, where $X$ denotes the HE-encrypted ciphertext and $W_c$ is a weight matrix:
\vspace*{-0.03in}
\begin{align}
Y &=  X \times W_c, 
\end{align}

\noindent\textbf{$\mathcal{S}$(LayerNorm):}
Referring to Formula~\ref{LayerNorm}, during secure inference, the LayerNorm function, which involves computing the reciprocal square root to obtain $\sigma$, is typically implemented using linear approximation methods \cite{hao2022iron,pang2023bolt}.

\vspace*{0.03in}
\noindent\textbf{$\mathcal{S}$(ln3):}
Similar to $\mathcal{S}$(ln2), this operation involves HE multiplication in the form of $\text{CT} \times \text{Weight}$, where $X$ denotes the HE-encrypted ciphertext and $W_1$ is a weight matrix:
\begin{align}
Y &=  X \times W_1, 
\end{align}

\noindent\textbf{$\mathcal{S}$(GELU):}
Referring to Formula~\ref{GELU}, in secure inference, the GELU function is usually approximated using a multi-degree polynomial \cite{hao2022iron,pang2023bolt,lu2023bumblebee}.

\vspace*{0.03in}
\noindent\textbf{$\mathcal{S}$(ln4):}
Similar to $\mathcal{S}$(ln3). Here, $X$ denotes the HE-encrypted ciphertext and $W_2$ is a weight matrix:
\begin{align}
Y &=  X \times W_2, 
\end{align}

\section{System Description}
\label{SecDTD_sec3}

This section sequentially introduces SecDTD, a Dynamic Token Drop method designed for Secure Transformer Inference. Subsection 3.1 proposes a novel pre-Softmax scoring scheme that evaluates token importance independently of Softmax outputs, thereby enabling early-gain token dropping. This is followed by a corresponding security analysis. Subsection 3.2 presents a Fast Oblivious Median Selection protocol that efficiently and securely identifies the median of the score vector, supporting precise half-and-half token dropping; this subsection is also accompanied by a dedicated security analysis. Finally, building upon these techniques, Subsection 3.3 introduces our token drop methodology that maximizes the efficiency of token drop.

\subsection{Pre-Softmax Token Scoring, MCN}
\label{sec_MCN}

As the initial step of Token Drop, it is crucial to efficiently and accurately assess the importance of each input token to eliminate less important tokens while maintaining inference accuracy. Scoring strategies for this purpose can be broadly categorized into static and dynamic approaches. Static strategies generate a fixed importance score list for input word vectors by analyzing the training dataset and summarizing inherent patterns. In this method, the importance of each position in the input sequence remains constant, regardless of the current input sequence, and token drop is performed accordingly. In contrast, dynamic scoring strategies evaluate the importance of each token within the current input sequence, dynamically producing an  importance score list with regard to the specific input. This approach better reflects real-world inference scenarios, where user inputs can vary significantly, and the importance of a token at a given position varies depending on the input. Compared to static methods that rely on prior knowledge, dynamic scoring offers greater flexibility and responsiveness, making it more suitable for practical applications.

\begin{figure}[t]
\centering
\includegraphics[trim={0cm 0cm 0cm 0cm}, clip, scale=0.59]{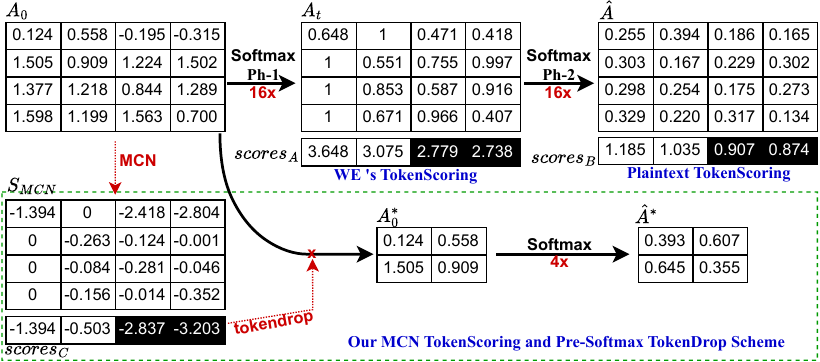}
\vspace*{-0.2in}
\caption{Comparison of Scoring Methods.} 
\label{Score_MCN}
\end{figure}

Existing dynamic scoring methods rely on the attention matrix $\hat{A} = \text{Softmax}(A_0)$, obtained through the Softmax operation, as illustrated in Figure~\ref{Score_MCN}, specifically in \textit{Plaintext TokenScoring}. In the self-attention mechanism of Transformers, each element $x_{ij}$ in the attention matrix $\hat{A}_{m \times m}$ represents the degree to which token $i$ attends to token $j$. During attention computation, token $i$ evaluates all tokens, including itself, to determine which ones to prioritize. The score $x_{ij}$ quantifies this level of attention.
A common approach to measuring token importance is to sum the values within a column of the attention matrix $\hat{A}_{m \times m}$, effectively aggregating all $x_{ij}$ values for a fixed $j$. This aggregated score represents the total attention token $j$ receives from all other tokens in the sequence. A higher score indicates that token $j$ is attended to by multiple tokens, suggesting higher importance, while a lower score implies reduced relevance.
For multi-head self-attention, importance scores are further aggregated across multiple attention heads by summing the corresponding attention matrices, providing a more comprehensive evaluation of token importance.

However, a clear limitation of existing methods is that the token importance score used for token drop depends on the Softmax output. This reliance means that Softmax must be completed before token drop occurs, negating any potential efficiency gains. 
While this issue is often overlooked in plaintext inference, where the Softmax computation accounts for a negligible portion of runtime, it becomes a significant and pressing concern when transitioning to secure inference, where the cost of Softmax increases sharply. To address this problem, we propose a scoring method that eliminates the need for Softmax. Our approach accurately evaluates token importance while introducing minimal additional overhead.

First, we observe that the raw attention matrix $A_0 = Q \times K^T$ contains the unprocessed attention information for each token. However, these independent attention values, having not undergone Softmax normalization, cannot be directly integrated to reliably reflect token importance. Softmax, as a normalization operation, consists of two core components—exponential transformation (Ph-1) and summation-based normalization (Ph-2)—as described in Formula~\ref{softmax_2ph}. Its primary functions are nonlinear scaling and normalization. Notably, in the Multi-Party Computation (MPC) environment, the exponential operation is particularly costly, accounting for over 80\% of the Softmax runtime.
\textit{Through our evaluation, we find that while the exponential is necessary for Softmax function, its precision is excessive for scoring token importance.} Based on experimental verification, we propose a Token Importance Scoring method optimized for MPC environments—\textbf{Max-Centric Normalization (MCN)}. MCN normalizes each element within a set by calculating its deviation from the maximum value and applying an additional scaling factor to reduce its magnitude. The formal definition of MCN is provided as follows:
\begin{align}
\label{MCN_formula1}
MCN(x)_{ij} &=  \frac{x_{ij} - \max_i}{\max_i^n}, \\
&= (\frac{x_{ij} - \max_i}{\max_i})\times (\frac{1}{\max_i})^{n-1} \nonumber
\end{align}
\noindent where $\max_{i}$ is the maximum value of the $i$-th row $[x_{i0}, x_{i1}, ..., x_{im}]$, with $i \in [m]$ and $j\in [m]$. The parameter $n$ can be adjusted according to the scale of values, and for the dataset used in this work, we set $n=2$.

\vspace*{0.1in}

\noindent\textbf{Interpretation and Benefits:}
\vspace*{-0.02in}
\noindent
\begin{itemize}[leftmargin=0in, itemindent=0.15in]
    \item \textbf{Deviation from Maximum:} The term $\frac{x - \max}{\max}$ captures the relative deviation of each element from the maximum, bringing values closer to zero as they approach the maximum.   
    \item \textbf{Additional Scaling Effect:} The factor $\left(\frac{1}{\max}\right)^{n-1}$ serves as a scaling mechanism, further reducing the range of values. This is particularly beneficial in datasets with large or extreme values, as it compresses values closer to zero and minimizes the influence of large values and outliers.
    \item \textbf{Efficiency in MPC Environments:} MCN is designed with two core components: $(x - \max)$ and $(\max^n)$. Since the \bm{$\max$} value, as a prerequisite for the Softmax calculation, is already computed, it does not incur additional overhead. In this way, the former can be executed locally ``for free" in MPC environments with Additive Secret Sharing. The latter can be simplified into a reciprocal multiplication operation, which is negligible compared to the overall inference cost.

\end{itemize}

Overall, the design principle of MCN is to fulfill the required functionality while minimizing additional overhead.
MCN provides a refined method for data normalization, effectively balancing deviation with controlled magnitude adjustment through scaling. This facilitates token drop before the Softmax operation, significantly enhancing the benefit of token drop. As shown in  Figure~\ref{Score_MCN}, while existing approaches  require conducting Softmax on 16 values, our method reduces this to only 4 values.

\begin{figure}[t]
\centering
\includegraphics[trim={0cm 0cm 0cm 0cm}, clip, scale=0.5]{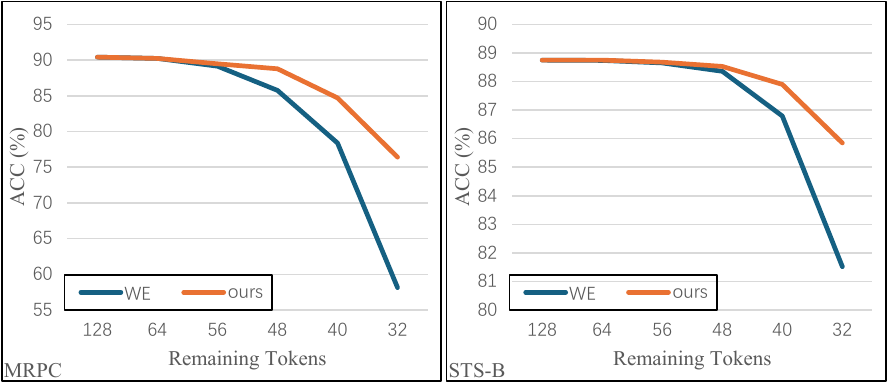}
\vspace*{-0.1in}
\caption{Token Drop Applied to the 1st BERT Layer, Reducing Tokens from 128 to 64, 56, 48, 40, and 32.}
\label{Test-1}
\end{figure}

Moreover, the token scores generated by MCN enable a more precise evaluation of token importance. The existing scoring method, W.E., possibly for the sake of simplification, utilizes the intermediate result of Softmax Ph-1. However, without the Softmax Ph-2 component, W.E.’s scoring process may be influenced by extreme values due to insufficient normalization. In contrast, the proposed method, incorporating both normalization and scaling function, is more robust in scenarios involving high dynamic range data with extreme values, which commonly occur in extreme compression scenarios.
To simulate these scenarios and evaluate the robustness of each scoring scheme, we conducted two sets of stress test suites on the MRPC and STS-B datasets with an input sequence length of 128:

\noindent\textbf{Test-1: Single-Layer Gradual Token Drop Challenge.} As depicted in Figure~\ref{Test-1}, token drop is applied to the first layer of BERT, gradually reducing the number of tokens from 128 to 64, 56, 48, 40, and finally 32.

\begin{figure}[t]
\centering
\includegraphics[trim={0cm 0cm 0cm 0cm}, clip, scale=0.5]{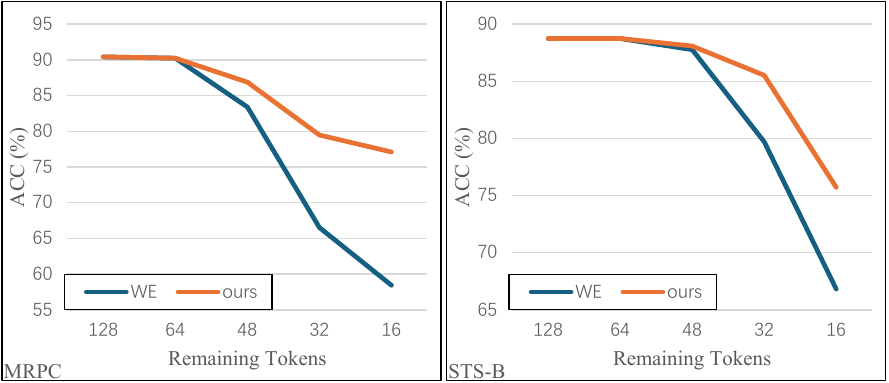}
\vspace*{-0.1in}
\caption{Token Drop is Progressively Applied from the 1st to 4th Layer, Reducing the Tokens from 128 to 64, 48, 32, and 16 at each Layer.}
\vspace*{-0.1in}
\label{Test-2}
\end{figure}

\noindent\textbf{Test-2: Cumulative Layers Token Drop Challenge.} As shown in Figure~\ref{Test-2}, token drop is applied progressively from the 1st to 4th layer, reducing the number of tokens from 128 to 64, 48, 32, and 16 at each respective layer.

In both scenarios, when token drop is relatively mild, both methods perform well. However, under deep token drop conditions, W.E. experiences a rapid collapse in accuracy, whereas MCN, despite some accuracy loss, remains significantly more stable. This highlights the superior robustness of the MCN method in handling extreme values.
Notably, even with a potential accuracy drop, the W.E. approach does not achieve simplification very well. This is because, in the Softmax operation, Ph-1 remains the dominant computational component, while the omitted Softmax Ph-2 accounts for only about 20\% of the total Softmax overhead in an MPC environment.

Finally, MCN offers another advantage through its contribution to the subsequent median searching process. The score vector processed using MCN exhibits a well-structured distribution, characterized by a smooth and gradually decreasing trend, as shown in Figure~\ref{MCN_avg_median}. This property ensures that the \textbf{average} closely approximates the \textbf{median}, thereby accelerating the average-as-pivot median searching process. Further details will be discussed in the following section.

\vspace*{0.1in}
\noindent\textbf{Security Analysis of MCN.}
\label{MCN_security}
The security analysis of the proposed MCN is straightforward. As shown in Formula~\ref{MCN_formula1}, MCN reuses several existing cryptographic protocols for basic computations, including multiplication, addition, reciprocal, and negation, all of which can be securely realized using OT in a MPC environment. As a result, MCN is considered secure.

\subsection{Fast Oblivious Median Selection, OMSel}
\label{OMSel}

Once the importance scores of all tokens are obtained, the next crucial step is determining which tokens to be dropped based on these scores.
W.E.\cite{pang2023bolt} adopts a half-and-half token drop approach, removing half of the current tokens at each interaction by discarding those with scores below the median. CipherPrune \cite{zhang2025cipherprune} uses a static threshold learned through fine-tuning, dropping tokens with scores lower than this fixed threshold. As discussed earlier, static thresholding not only incurs the cost of fine-tuning to determine the optimal threshold but also suffers from poor packing efficiency when used with packed HE primitives, as it lacks precise control over the number of retained tokens. More critically, it may introduce potential information leakage.
Similar to W.E., we use the half-and-half token drop strategy. However, we propose a novel protocol that significantly enhances the median searching process.
The existing W.E. scheme relies on Bitonic Sort. While it benefits from parallel processing, it still incurs high complexity in MPC environments. Therefore, an optimal design must prioritize two key objectives: reducing complexity and ensuring security.

\begin{figure}[h]
\centering
\includegraphics[trim={0cm 0cm 0cm 0cm}, clip, scale=0.98]{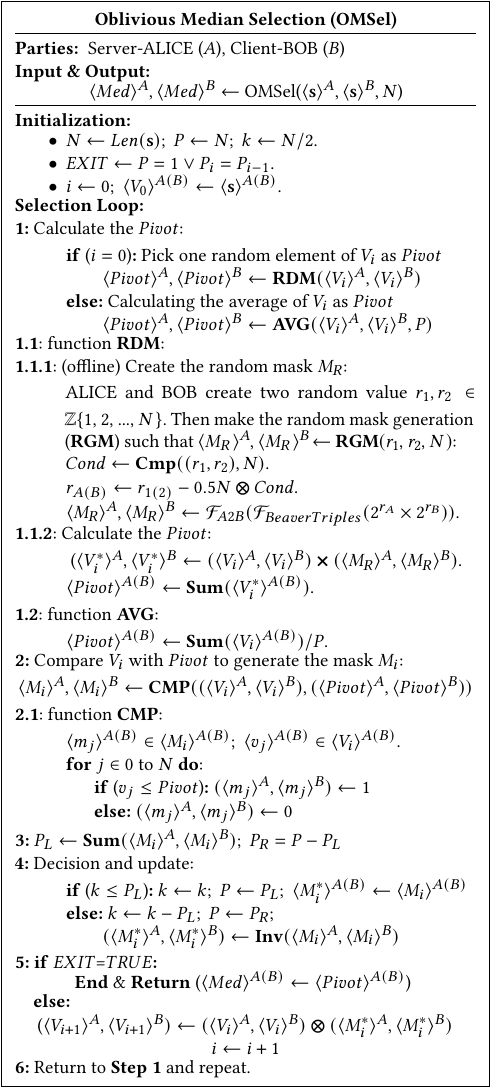}
\vspace*{-0.1in}
\caption{Protocol Oblivious Median Selection (OMSel).}
\label{Protocol_OMSel}
\end{figure}

To address these challenges, we propose a secure and fast median search method, \textbf{Oblivious Median Selection (OMSel)}, as illustrated in Figure~\ref{Protocol_OMSel}. OMSel efficiently identifies the median while maintaining security, ensuring obliviousness and randomness. Additionally, Figure~\ref{Toy_example_OMSel} in Appendix provides a toy example that demonstrates the OMSel process in practice.
In the following discussion, we introduce the design principles and advantages of OMSel.

In MPC environments, complexity is primarily determined by two costly operations: \textbf{\#Cmp} and \textbf{\#Mux (denoted as \bm{$\otimes$})}\cite{nishide2007cmp,catrina2010secure}.
\#Cmp, or Secure Comparison, is a protocol that allows two or more parties to compare numerical values without revealing their inputs, commonly applied in the Millionaire’s Problem.
\#Mux, or Secure Multiplexer, is used to select an output from multiple data paths based on a selection bit. In secure computation, this is typically implemented as a Secure Select operation.
In secure inference frameworks, these operations are generally executed using Secret Sharing and OT. Their overhead dominates the median search process \cite{demmler2015aby}. Therefore, evaluating the number of \#Cmp and \#Mux operations is essential for assessing the efficiency of different median search designs.
For example, Bitonic Sort requires 4608 \#Cmp and 9216 \#Mux operations to find the median of a 256-element vector. This introduces considerable overhead, taking over 18 seconds under a 200 Mbps, 80 ms network condition.

Bitonic Sort is excessive for the median-searching task, as it sorts the entire score array and introduces unnecessary computational overhead. \textit{Given that the objective is merely to identify a single element—the median—this inefficiency motivates the adoption of more targeted approaches, drawing inspiration from selection algorithms.}
A classic selection algorithm is QuickSelect, which efficiently finds the $k$-th smallest element in an array. It is based on the partitioning strategy of QuickSort but only recurses on a portion of the data, making it more efficient than full sorting in terms of time complexity.
However, QuickSelect is not \textbf{oblivious}, meaning it does not hide access patterns. Its partitioning process, which progressively narrows the recursion range, directly exposes data comparisons and swaps, leading to potential information leakage. In an MPC environment, the non-deterministic \textit{access pattern} of QuickSelect could reveal sensitive information about the data distribution \cite{beame2015finding,bogdanov2014practical}.
To address this, we leverage the advantages of partitioning while ensuring obliviousness. Instead of directly discarding the ``eliminated" elements, we use a secure mask to track the status of each element (whether it is eliminated or not), maintaining a consistent access pattern throughout the process. This creates a \textbf{pseudo-partitioning} process, where elements are not physically moved or removed but are instead logically masked. By doing so, we retain the efficiency benefits of partitioning while preventing information leakage, ensuring a fully oblivious execution.

Next, as a selection protocol that employs a partitioning strategy, choosing an \textbf{effective pivot} is crucial. A well-chosen pivot can significantly enhance the efficiency of the selection process. Furthermore, since this protocol operates within a secure inference framework, the pivot must remain confidential throughout the entire process to prevent any potential information leakage. To achieve this, we strategically select the average value of the score vector as the pivot for the following reasons:
\begin{itemize}[leftmargin=0in, itemindent=0.15in]
\item \textbf{Security and Cost Efficiency:} The averaging operation involves summation and division by the total count. Summation can be performed locally in a secure and efficient manner within MPC environments using Additive Secret Sharing. And the division by total count $N$ is secure because $N$ remains constant, as our pseudo-partitioning strategy retains all elements of the score vector.
\item \textbf{Performance Enhancement:} Selecting a pivot close to the median is desirable for median selection, and the average value serves as an effective approximation. As discussed in the previous section, MPC ensures a gradually decreasing value distribution, which allows the average to closely approximate the median, as illustrated in Figure~\ref{MCN_avg_median}. Consequently, using the average as the pivot accelerates convergence in median selection.
\end{itemize}
\vspace*{-0.02in}


These design choices integrate seamlessly to enhance the overall performance of our protocol.
However, relying entirely on the average as the pivot introduces a potential risk. Unlike a selected pivot, the average is computed rather than chosen. For a score vector with identical values, the computed average will always be the same. This means a client could craft a specifically designed input that results in an identical score vector and, consequently, the same average as the pivot. While this does not expose individual element values or access patterns, a \textit{consistent operational pattern} may still pose a security risk against \textbf{response attacks} from a malicious client \cite{tramer2016stealing,shokri2017membership}.
Although malicious clients are outside the scope of our threat model, which assumes a semi-honest setting, and response attacks can be mitigated through additional techniques such as differential privacy \cite{dwork2014algorithmic}, we take a proactive approach to eliminate this risk at its source. Our solution introduces \textbf{randomness} into the pivot selection process. Specifically, in the initial round, we use a randomly selected element as the pivot (\textbf{RDM}).
To securely set this random pivot, during the offline initialization phase, the two-party computation (2PC) participants—the server and the client—each generates its own random number, $r_1$ and $r_2$, respectively. These values are used to jointly determine a secret index pointing to a random position within the score vector. Finally, a corresponding random mask is generated using Beaver Triples, with each party holding a share. During the online inference phase, this random mask enables the efficient selection of an element from the score vector as the pivot, thereby introducing randomness and fundamentally eliminating the risk of response attacks. (\textit{Note: In the toy example shown in Figure~\ref{Toy_example_OMSel}, RDM is omitted for simplicity of presentation.})

\vspace*{0.1in}
\noindent\textbf{Security Analysis of OMSel.}
\label{MCN_security}
We now analyze the security of the proposed OMSel.
In the semi-honest setting, the goal is to prevent the server from learning the client’s private input data, and conversely, block the client from accessing the server’s model parameters.
First, all operations are performed using established MPC primitives, including multiplication, addition, secure comparison, and multiplexer functions. This ensures that the actual token score values and masks remain securely hidden from both the client and the server.
Second, to maintain obliviousness, we adopt a pseudo-partitioning strategy in the median search. This ensures that all token scores remain logically engaged in the search process—no real partition or elimination is conducted. As a result, neither party can infer any information about the relative ranking between tokens (e.g., whether token A is more significant than token B) or any intermediate decisions regarding retention or removal.
Third, randomness is introduced through randomized pivot selection. This prevents the formation of deterministic access patterns and blocks the client from crafting special inputs that could infer model characteristics based on consistent operational traces.
Therefore, under the semi-honest model, neither party can actively extract the other’s secret.

We further examine the potential for accidental leakage. For instance, could a weak input—such as a sequence where all tokens are identical—lead to identical token scores and thus a deterministic exit pattern that reveals input distribution?
The answer is NO. Even if the input tokens are identical, Transformer models incorporate positional encoding, making each token’s representation unique. Additionally, the attention mechanism is a highly non-linear process; it does not map identical inputs to identical scores. Empirically, we have never observed two tokens with exactly identical significance scores in any evaluation.

Finally, our design is consistent with prior work under the standard semi-honest assumption~\cite{pang2023bolt,zhang2025cipherprune,lu2023bumblebee,rathee2020cryptflow2,279898}. However, we also aim to enhance security beyond this assumption. As noted earlier, some existing token drop methods pose a risk of information leakage due to the one-to-one mapping between specific inputs and token drop decisions. This behavior can act as a ``fingerprint", introducing potential vulnerabilities under stronger threat models such as adversarial settings with more powerful attackers. To eliminate this risk, the proposed SecDTD incorporates two key design choices. First, the same input does not consistently lead to a fixed token drop pattern, owing to the intentional introduction of randomness, as previously discussed. Second, SecDTD enforces consistent drop decisions across different inputs for the same task—that is, the same number of tokens is dropped regardless of input variation. While this uniformity may slightly reduce flexibility in achieving optimal speedups, the resulting security benefits are significant. It completely removes the one-to-one mapping between inputs and drop patterns, thereby mitigating the fingerprinting threat.

\begin{table*}[t]
\vspace*{-0.1in}\caption{Practical Complexity Comparison: OMSel vs. Bitonic Sort Across Real-World Datasets.}\vspace*{-0.1in}
\label{OMSel_exp}
\centering
\begin{tabular}{cr|c|ccccc|cc}
\Xhline{1.5pt}
\multicolumn{2}{c|}{\textbf{Dataset}}                             & \textbf{CoLA}                        & \textbf{SST-2}                        & \textbf{MRPC}                        & \textbf{STS-B}                        & \textbf{QNLI}                        & \textbf{QQP}                          & \textbf{MNLI}                         & \textbf{RTE}                         \\ \hline
\multicolumn{2}{c|}{\textbf{Input Sequence Length}}                    & 64                                   & 128                                   & 128                                  & 128                                   & 128                                  & 128                                   & 256                                   & 256                                  \\ \Xhline{0.8pt}
\multicolumn{1}{c|}{}                                 & \# Cmp    & { 672}           & { 1792}           & { 1792}          & { 1792}           & { 1792}          & { 1792}           & { 4608}           & { 4608}          \\
\multicolumn{1}{c|}{\multirow{-2}{*}{Bitonic Sort}}                                & \# Mux & { 1344}          & { 3584}           & { 3584}          & { 3584}           & { 3584}          & { 3584}           & { 9216}           & { 9216}          \\ \Xhline{0.8pt}
\multicolumn{1}{c|}{}                                 & \# Cmp    & { 88}            & { 176}            & { 404}           & { 217}            & { 382}           & { 210}            & { 339}            & { 598}           \\
\multicolumn{1}{c|}{\multirow{-2}{*}{\textbf{OMSel}}}                             & \# Mux & { 88}            & { 176}            & { 404}           & { 217}            & { 382}           & { 210}            & { 339}            & { 598}           \\ \Xhline{0.8pt}
\multicolumn{2}{c|}{\textbf{Aggregated Speedup}}                             &  $\bm{9.6\times}$ &  $\bm{12.8\times}$ &  $\bm{5.5\times}$ &  $\bm{10.4\times}$ & $\bm{5.9\times}$ & $\bm{10.7\times}$ & $\bm{16.9\times}$ & $\bm{9.6\times}$ \\ \Xhline{1.5pt}
\end{tabular}
\vspace*{-0.1in}
\end{table*}

\vspace*{-0.05in}
\begin{table}[h]
\centering
\caption{Complexity Comparison: OMSel vs. Bitonic Sort.}
\label{Complexity_OMSel_Bitonic}
\vspace*{-0.05in}
\begin{tabular}{c|cc}
\hline
\textbf{Method}     & Bitonic Sort & OMSel \\ \hline
\textbf{\# Cmp} & $N\log^2N$       & $N\log N$ \\ \hline
\textbf{\# Mux} & $2N\log^2N$       & $N\log N$ \\ \hline
\end{tabular}
\vspace*{-0.2in}
\end{table}

\vspace*{0.1in}
\noindent\textbf{Complexity Comparison.}
Table~\ref{Complexity_OMSel_Bitonic} presents the complexity of each method, specifically the number of \#Cmp and \#Mux operations, for Bitonic Sort used by W.E. and our proposed OMSel. It is evident that OMSel offers a  complexity advantage, which becomes even more pronounced as the size of the score array, i.e., the input sequence, increases. A detailed complexity analysis and proof can be found in the Appendix-\ref{Complexity_Analysis}.
Table~\ref{OMSel_exp} provides comprehensive evaluations, displaying the exact counts of \#Cmp and \#Mux operations, as well as the execution runtime across different dataset tasks. Due to the inherent randomness in the process, the reported performance of OMSel represents the average result over five evaluations.
Compared to Bitonic Sort, OMSel achieves up to \bm{$16.9\times$} acceleration, significantly reducing the execution runtime. This substantial improvement enhances the practicality of OMSel for processing long input sequences and supporting multiple rounds of token drop.

\subsection{Dynamic Token Drop for Secure and Efficient Transformers Inference, SecDTD}
\label{HE-Efficient-Network}

\vspace*{-0.05in}
With the introduction of \textbf{MCN}, which scores token importance prior to the Softmax operation, and \textbf{OMSel}, which quickly and securely identifies the median of the token score array to determine token removal by discarding those with scores below the median, the proposed \textbf{SecDTD} strategically advances the token drop operation to earlier stages of Transformer inference. This adjustment aligns with the unique cost distribution in ciphertext-based scenarios.
SecDTD not only optimizes $\mathcal{S}$(ln2), $\mathcal{S}$(ln3), and $\mathcal{S}$(ln4), but also extends the benefits of token dropping to earlier operations, including $\mathcal{S}$(Softmax), $\mathcal{S}$($\times$V), and even the initial computation stage of $I \times W_v = V$.

\begin{figure}[h]
\centering
\includegraphics[trim={0cm 0cm 0cm 0cm}, clip, scale=0.36]{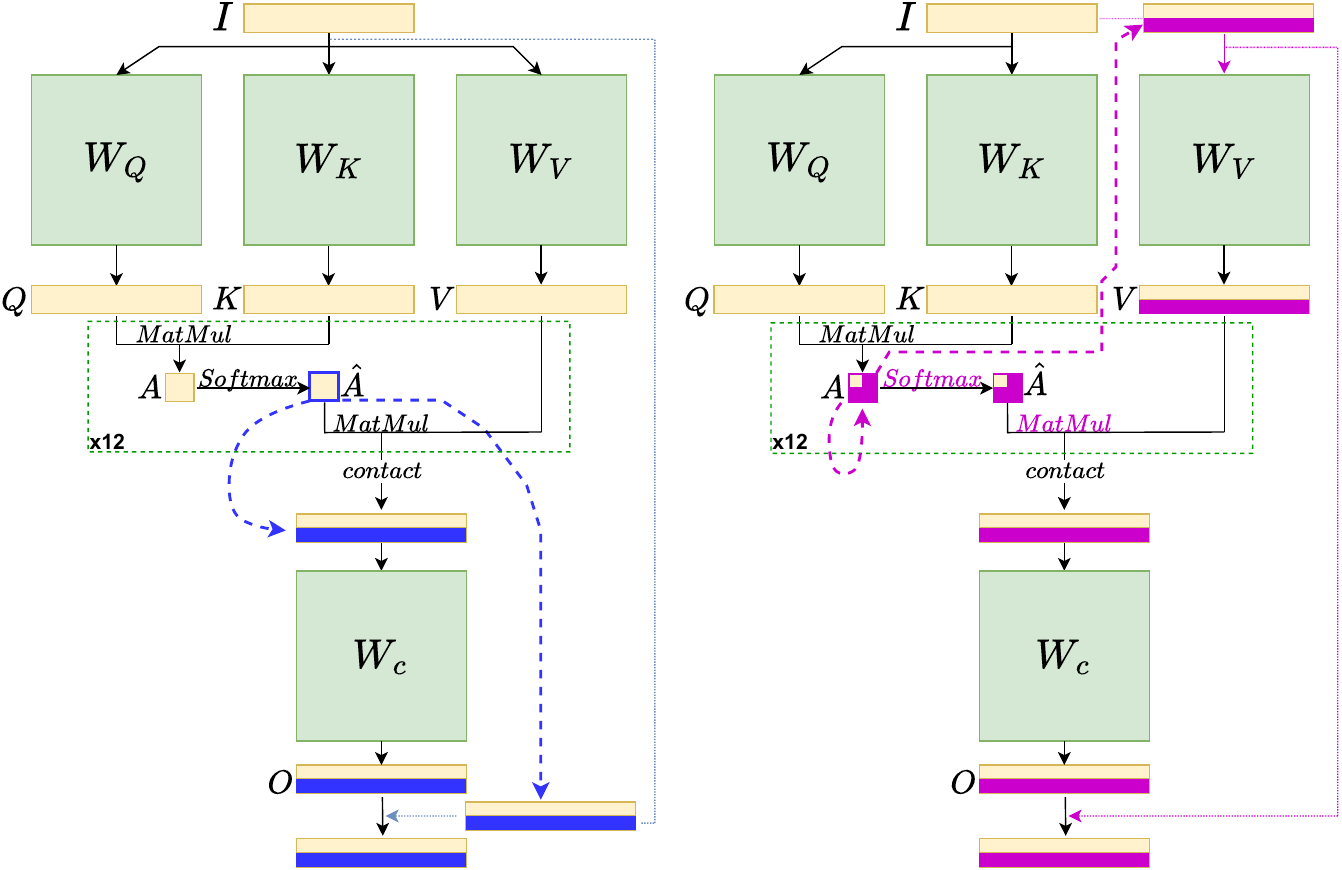}
\vspace*{-0.2in}
\caption{Comparison of Token Drop Schemes: Existing Approaches without Pre-Gain (Left) vs. SecDTD (Right). Highlight Differences in Where and How Token Drop (color-coded) is Applied, Demonstrating How SecDTD's Pre-softmax Design Benefits More Stages of Inference.}
\label{vsWE_bolt}
\end{figure}

\begin{figure}[t]
\centering
\includegraphics[trim={0cm 0cm 0cm 0cm}, clip, scale=0.5]{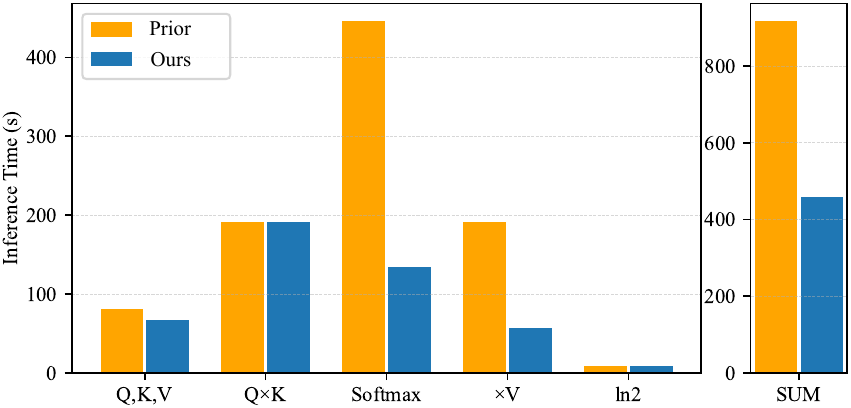}
\vspace*{-0.1in}
\caption{Runtime Comparison in BOLT framework.} 
\vspace*{-0.25in}
\label{SecDTDvsWE}
\end{figure}

\begin{figure}[t]
\centering
\includegraphics[trim={0cm 0cm 0cm 0cm}, clip, scale=0.5]{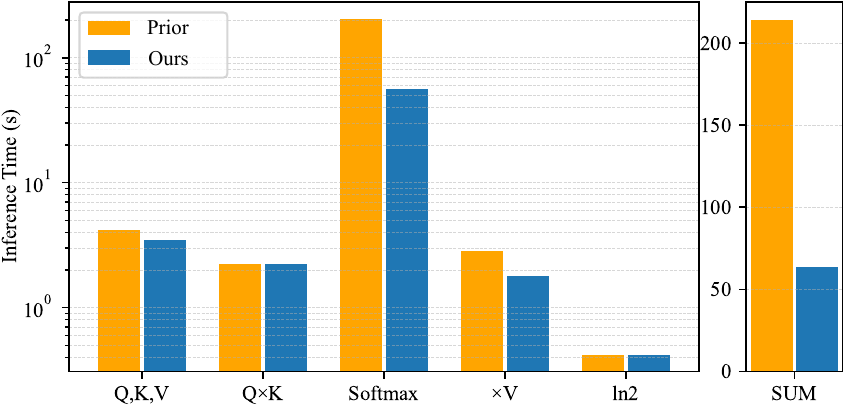}
\vspace*{-0.1in}
\caption{Runtime Comparison in BumbleBee framework.} 
\vspace*{-0.1in}
\label{SecDTDvsWE2}
\end{figure}

Figure~\ref{vsWE_bolt} illustrates the token drop mechanisms of both existing approaches \cite{pang2023bolt,zhang2025cipherprune} and the proposed SecDTD. On the left, prior methods adopt a token drop strategy similar to that used in plaintext settings, where the Softmax output, $\hat{A}$, is utilized to drop tokens from the input sequence at stage $\mathcal{S}$(ln2), formulated as $Y(X) = X \times W_c$, as well as from the residual connection sequence. This strategy provides computational benefits beginning at $\mathcal{S}$(ln2) and propagates them through subsequent stages.
On the right, the SecDTD token drop mechanism is illustrated. Instead of relying on the Softmax output, SecDTD directly uses the raw attention matrix ${A}$ from $\mathcal{S}$(Q$\times$K), supported by MCN and OMSel, to perform token drop on ${A}$ itself, resulting in a reduced input to Softmax. Additionally, such an earlier token drop is applied to the input of $V = I \times W_v$. Compared to existing methods, this design allows SecDTD to extend optimization to cover $I \times W_v = V$, Softmax, and $\times$V, maximizing performance gains.
As shown in Figure~\ref{SecDTDvsWE}, within the Secure Transformers Inference Framework BOLT, SecDTD achieves a significant reduction in inference time. With an input length of 256 tokens, SecDTD delivers a \textbf{2.04$\times$} speedup over existing methods. Under another widely adopted framework, BumbleBee, depicted in Figure~\ref{SecDTDvsWE2}, the performance gain becomes even more substantial, achieving a \textbf{3.37$\times$} inference time speedup. This improvement is primarily due to the higher proportion of Softmax in BumbleBee (note the logarithmic scale on the Y-axis).

Our dropping process follows the same procedure as prior token drop methods \cite{pang2023bolt,zhang2025cipherprune}, where tokens with scores below the median are iteratively moved to the end of the sequence using an OT-based oblivious swap, after which the tail tokens are discarded. The overhead introduced by this dropping process is minimal, accounting for less than 1\% of the overall inference time\cite{pang2023bolt,zhang2025cipherprune}.
The following section, Section~\ref{Implementation}, provides a detailed, layer-wise, end-to-end implementation of the token drop mechanism, demonstrating how the SecDTD method is applied to a real-world Transformer model, BERT-base.

\subsection{Security Guarantee}
\label{security}
SecDTD is built upon the state-of-the-art secure Transformer inference frameworks BOLT~\cite{pang2023bolt} and BumbleBee~\cite{lu2023bumblebee}, operating under the semi-honest adversary model. The security of linear computations is guaranteed by the semantic security of HE schemes, such as BFV~\cite{fan2012somewhat} and CKKS~\cite{cheon2017homomorphic}. For nonlinear operations, the framework employs Secret Sharing (SS)\cite{shamir1979share} and Oblivious Transfer (OT)\cite{brassard1986all}.
All operations introduced by SecDTD to enable MCN and OMSel—including addition, multiplication on secret shares, secure comparisons, and multiplexer functions—are securely executed in a two-party computation (2PC) setting between the Server and the Client, using Additive SS and OT. This protocol design aligns with the secure computation of nonlinear functions such as Softmax.
Based on the security analyses provided at the end of the MCN and OMSel subsections, the proposed SecDTD does not compromise the client’s input data, nor does it allow the client to access the server’s model parameters. Overall, SecDTD is considered secure under the semi-honest assumption. Attacks that fall outside this threat model are beyond the scope of this work.

\begin{figure*}[t]
\centering
\includegraphics[trim={0cm 0cm 0cm 0cm}, clip, scale=0.72]{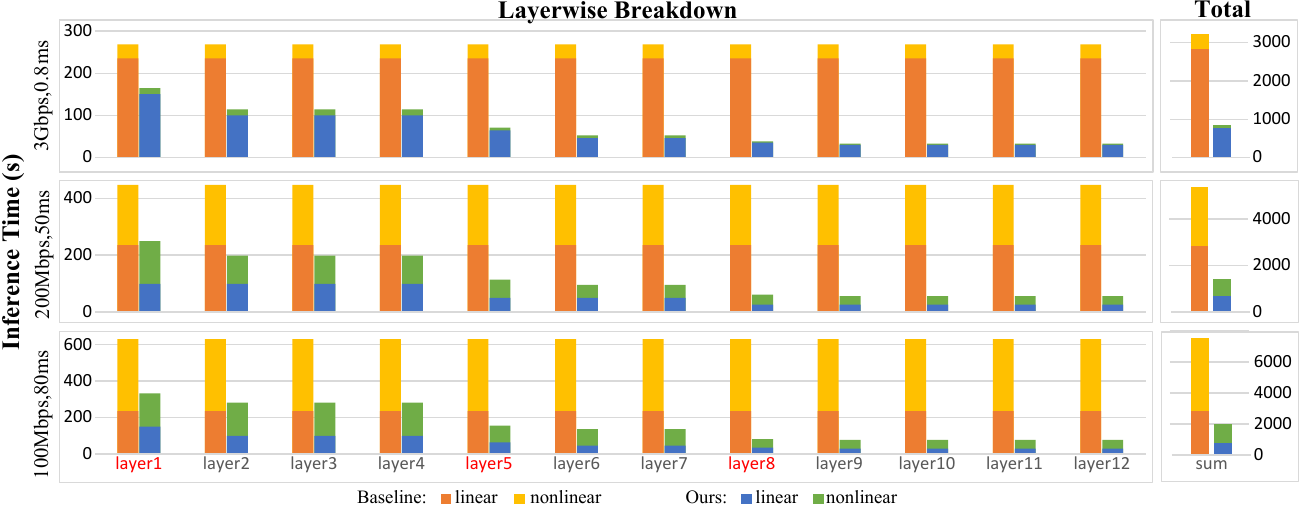}
\vspace*{-0.1in}
\caption{Implementation of BERT-Base on MRPC (under BOLT Framework).} 
\vspace*{-0.15in}
\label{layerwise}
\end{figure*}

\section{Evaluation}
\label{Evaluation}

\subsection{Implementation}
\label{Implementation}

For real-world implementation, we apply token drop on one Transformer layer using the proposed SecDTD mechanism, which involves MCN pre-Softmax scoring and OMSel median selection. A Transformer model, such as BERT, consists of multiple layers. By leveraging MCN's precise importance evaluation, token drop can be applied multiple times without compromising model accuracy.
SecDTD operates as a semi-flexible dynamic mechanism: the layers where token drop is applied are predetermined, while the specific tokens to be dropped are selected dynamically.
Figure~\ref{layerwise} illustrates the layer-wise, end-to-end implementation of the BERT-base model, which consists of 12 layers, on the MRPC dataset under the BOLT framework, where token drop is applied three times — specifically on layer-1, layer-5, and layer-8. 
These layers were manually chosen for simplicity; however, an offline auto-search could be conducted before implementation to identify the optimal layers for token drop, depending on the specific dataset and task \cite{goyal2020power,snoek2012practical}.

Notably, the evaluation is conducted without fine-tuning, as SecDTD dynamically scores each individual input sequence and performs precise token drop. This capability allows it to minimize inference overhead while maintaining accuracy to the greatest extent.
In the figure, the overall inference time is decomposed into a linear component, arising from linear computations primarily influenced by computational power, and a nonlinear component, resulting from nonlinear operations mainly affected by network conditions. Under high network speeds, linear computations dominate, whereas under low network conditions, the nonlinear component becomes the primary factor in the overall inference time. SecDTD achieves significant improvements in both linear and nonlinear components, thereby reducing the overall cost in all scenarios. This reduction is evident in both the layer-wise and end-to-end costs, as shown on the right side of the figure.
With a threefold token drop, the original input sequence is progressively reduced from 128 tokens to 64 tokens, then to 32 tokens, and finally to just 16 tokens. This substantial reduction significantly decreases inference time, achieving an overall speedup of more than $3.75\times$. Further evaluation results are presented in the following section.

\subsection{Experimental Setup}
\label{Experimental_Setup}

The experiments presented in this work are conducted on a testbed equipped with AMD 3995 64-core CPU. To simulate diverse network conditions, we utilized Linux Traffic Control. Specifically, evaluations were performed under three scenarios: \textbf{LAN} (3 Gbps, 0.8 ms), \textbf{WAN} (200 Mbps, 50 ms), and \textbf{Mobile} (100 Mbps, 80 ms). These configurations enable performance evaluations under varying bandwidths and latencies, aligning with real-world network conditions~\cite{speedtest}.

We implement SecDTD on the \textbf{BOLT framework} \cite{pang2023bolt}, utilizing the Secure and Correct Inference (SCI) library from EzPC \cite{EzPC} and the SEAL \cite{sealcrypto} library, as well as on the \textbf{BumbleBee framework} \cite{lu2023bumblebee}, which is built on top of the SPU library \cite{spu}, to evaluate its cost reduction performance. On each framework, we follow their default settings and parameters. Our objective is to evaluate the performance of SecDTD within each framework, specifically its speedup relative to the baseline. \textit{Notably, in experiments, the BOLT framework is configured with the basic number of threads, 2 threads, to ensure precise measurement, whereas the BumbleBee framework, configured by the SPU library, incorporates extensive optimizations and utilizes multi-threading}. Therefore, the focus should be on relative ratios and speedup factors in each experiment rather than absolute values or unit differences across frameworks.

\begin{table*}[t]
\centering
\vspace*{-0.0in}\caption{SecDTD End-to-End Performance under BOLT Framework without Fine-tuning.}\vspace*{-0.1in}
\label{BOLT_exp}
\begin{tabular}{ccr|cccccccc}
\Xhline{1.5pt}
\multicolumn{3}{c|}{\textbf{Dataset}}                                                                                                                                          & \textbf{CoLA} & \textbf{SST-2} & \textbf{MRPC} & \textbf{STS-B} & \textbf{QNLI} & \textbf{QQP} & \textbf{MNLI} & \textbf{RTE}  \\ \hline
\multicolumn{3}{c|}{\textbf{TASK}}                                                                                                                                             & ACPT          & SENT           & PARA          & SIM            & QA/NLI        & SIM          & NLI           & NLI                                      \\ \hline
\multicolumn{2}{c|}{\multirow{2}{*}{\textbf{ACC(\%)}}}                                                                                                              & Baseline & 58.80         & 91.74          & 90.53         & 87.97          & 91.25         & 87.68        & 84.69         & 68.23                                  \\
\multicolumn{2}{c|}{}                                                                                                                                               & ours     & 58.06         & 92.78          & 90.57         & 87.87          & 90.48         & 86.75        & 83.76         & 67.87                                    \\ \Xhline{0.8pt}
\multicolumn{1}{c|}{\multirow{3}{*}{\textbf{LAN}}}    & \multicolumn{1}{c|}{\multirow{2}{*}{\textbf{\begin{tabular}[c]{@{}c@{}}Inference\\ Time (s)\end{tabular}}}} & Baseline & 1370.46       & 3225.60        & 3225.60       & 3225.60        & 3225.60       & 3225.60      & 8464.14       & 8464.14                              \\
\multicolumn{1}{c|}{}                                 & \multicolumn{1}{c|}{}                                                                                       & ours     & 518.37        & 893.78         & 913.45        & 916.02         & 1316.63       & 886.03       & 1891.44       & 2355.26                               \\ \cline{2-11} 
\multicolumn{1}{c|}{} &
\multicolumn{2}{c|}{\textbf{Speedup}} &
$\bm{2.64\times}$ &
$\bm{3.61\times}$ &
$\bm{3.53\times}$ &
$\bm{3.52\times}$ &
$\bm{2.45\times}$ &
$\bm{3.64\times}$ &
$\bm{4.47\times}$ &
$\bm{3.59\times}$ \\
\Xhline{0.8pt}
\multicolumn{1}{c|}{\multirow{3}{*}{\textbf{WAN}}}    & \multicolumn{1}{c|}{\multirow{2}{*}{\textbf{\begin{tabular}[c]{@{}c@{}}Inference\\ Time (s)\end{tabular}}}} & Baseline & 2377.48       & 5368.23        & 5368.23       & 5368.23        & 5368.23       & 5368.23      & 13475.92      & 13475.92                             \\
\multicolumn{1}{c|}{}                                 & \multicolumn{1}{c|}{}                                                                                       & ours     & 887.52        & 1509.03        & 1549.26       & 1533.24        & 2241.65       & 1519.39      & 3088.78       & 3837.30                              \\ \cline{2-11} 
\multicolumn{1}{c|}{} &
\multicolumn{2}{c|}{\textbf{Speedup}} &
$\bm{2.68\times}$ &
$\bm{3.56\times}$ &
$\bm{3.47\times}$ &
$\bm{3.50\times}$ &
$\bm{2.39\times}$ &
$\bm{3.53\times}$ &
$\bm{4.36\times}$ &
$\bm{3.51\times}$ \\
\Xhline{0.8pt}
\multicolumn{1}{c|}{\multirow{3}{*}{\textbf{MOBILE}}} & \multicolumn{1}{c|}{\multirow{2}{*}{\textbf{\begin{tabular}[c]{@{}c@{}}Inference\\ Time (s)\end{tabular}}}} & Baseline & 3393.21       & 7564.60        & 7564.60       & 7564.60        & 7564.60       & 7564.60      & 18638.96      & 18638.96                               \\
\multicolumn{1}{c|}{}                                 & \multicolumn{1}{c|}{}                                                                                       & ours     & 1250.05       & 2122.99        & 2183.75       & 2149.83        & 3172.64       & 2150.96      & 4301.92       & 5346.08                              \\ \cline{2-11} 
\multicolumn{1}{c|}{} &
\multicolumn{2}{c|}{\textbf{Speedup}} &
$\bm{2.71\times}$ &
$\bm{3.56\times}$ &
$\bm{3.46\times}$ &
$\bm{3.52\times}$ &
$\bm{2.38\times}$ &
$\bm{3.52\times}$ &
$\bm{4.33\times}$ &
$\bm{3.49\times}$ \\
\Xhline{1.5pt}
\end{tabular}
\vspace*{-0.1in}
\end{table*}

To demonstrate the effectiveness of our SecDTD token drop mechanism, we evaluate it on eight datasets from the GLUE benchmark \cite{wang2018glue}, a widely used standard for assessing BERT’s performance. These datasets cover a range of linguistic tasks, including the Acceptability (ACPT) task: \textbf{CoLA}, the Sentiment (SENT) task: \textbf{SST-2}, the Paraphrase (PARA) task: \textbf{MRPC}, the Similarity (SIM) tasks: \textbf{STS-B} and \textbf{QQP}, the QA/NLI task: \textbf{QNLI}, and the NLI tasks: \textbf{MNLI} and \textbf{RTE}. These tasks involve varying input sequence lengths, with token counts ranging from 64 to 128 and 256 tokens. We employ trained models sourced from publicly available repositories \cite{cola,mrpc,qqp}.

\subsection{SecDTD Performance on BOLT Framework}
\label{SecDTD_BOLT}

\subsubsection{End-to-End Performance}
\label{End-to-End_Performance}

We present the end-to-end performance of SecDTD on eight datasets under three network settings in Table~\ref{BOLT_exp}. Without any fine-tuning, SecDTD achieves substantial reductions in inference time, with speedups of up to \bm{$4.47\times$}, all while preserving model accuracy.
Although the reported inference time in the table may appear high, it is important to note that experiments were conducted using 2 out of 128 CPU threads to ensure precise measurement and fair comparison with the baseline. In real-world deployments, multi-threading or multi-CPU setups are standard practice. For instance, with full utilization of dual CPUs, the MNLI task, originally requiring 70.5s, can be reduced to an estimated 15.7s with the $4.47\times$ speedup, underscoring the practicality of our method in real-world applications.
Due to its effectiveness across all stages of secure Transformer inference, including both linear and non-linear components, SecDTD consistently delivers strong performance across datasets of varying scales, task types, and network conditions, without relying on any specific scenario for optimal results.

\begin{figure}[t]
\centering
\includegraphics[trim={0cm 0cm 0cm 0cm}, clip, scale=0.74]{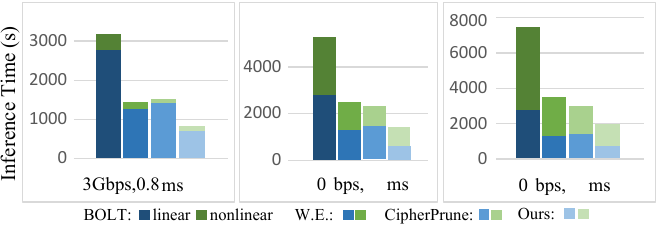}
\vspace*{-0.25in}
\caption{End-to-End Performance Comparison.} 
\vspace*{-0.15in}
\label{BOLT_WE_Cipher_ours}
\end{figure}

We further compare our proposed SecDTD with existing token drop methods, W.E.\cite{pang2023bolt} and CipherPrune\cite{zhang2025cipherprune}, as shown in Table~\ref{VS_BOLT} and Figure~\ref{BOLT_WE_Cipher_ours}. 
For a fair comparison, we report results on the \textbf{SST-2} dataset, which is originally used by all BOLT, W.E., and CipherPrune, under three different network settings.
The table also highlights five key properties: \textbf{FT-Free} (whether the method requires fine-tuning), \textbf{Sort-Free} (whether full sorting is needed), \textbf{Exact-Drop} (whether the method allows exact control over the number of dropped tokens), \textbf{Pre-Gain} (whether pre-Softmax benefits are available), and \textbf{Trace-Risk} (whether the method is vulnerable to potential leakage due to one-to-one mappings between specific inputs and token drop decisions).
SecDTD achieves the best inference latency across all network conditions while maintaining a high accuracy of 92.78\%, owing to its advantages across all five key properties.


\begin{table}[t]
\centering
\vspace*{-0.1in}\caption{End-to-End Performance Comparison}\vspace*{-0.1in}
\label{VS_BOLT}
\resizebox{82.8mm}{!}{
\begin{tabular}{c|c|ccc|ccccc}
\Xhline{1.5pt}
\multicolumn{1}{l|}{\multirow{2}{*}{\textbf{Method}}} & \multirow{2}{*}{\textbf{Acc.}} & \multicolumn{3}{c|}{\textbf{Inference Time (s)}}                                                           & \multirow{2}{*}{\textbf{\begin{tabular}[c]{@{}c@{}}FT\\Free\end{tabular}}} & \multirow{2}{*}{\textbf{\begin{tabular}[c]{@{}c@{}}Sort\\Free\end{tabular}}} & \multirow{2}{*}{\textbf{\begin{tabular}[c]{@{}c@{}}Exact\\Drop\end{tabular}}} & \multirow{2}{*}{\textbf{\begin{tabular}[c]{@{}c@{}}Pre\\Gain\end{tabular}}} & \multirow{2}{*}{\textbf{\begin{tabular}[c]{@{}c@{}}Trace\\Risk\end{tabular}}}\\
\multicolumn{1}{l|}{}                                 &                                & \multicolumn{1}{c}{\textbf{LAN}} & \multicolumn{1}{c}{\textbf{WAN}} & \multicolumn{1}{c|}{\textbf{Mobile}} &                                                                                 &                                                                                  &                                                                                      \\ \hline
BOLT                                                  & 91.74                          & 3225.60                          & 5368.23                          & 7564.60                              & $\cdot$                                                                               & $\cdot$                                                                                 & $\cdot$                              & $\cdot$                            & $\cdot$                             \\
W.E.                                                  & 92.78                          & 1473.89                          & 2522.73                          & 3581.81                              & $\checkmark$                                                                               & $\cdot$                                                                                & $\checkmark$                                          & $\cdot$                          & $\cdot$                   \\
\cite{zhang2025cipherprune}                                                 & 92.75                          & 1547.24                          & 2274.21                          & 3003.83                              & $\cdot$                                                                               & $\checkmark$                                                                                & $\cdot$                                                     & $\cdot$                          & \faExclamationTriangle    \\ \hline
\textbf{Ours}                                         & 92.78                          & 893.78                           & 1509.03                          & 2122.99                              & $\checkmark$                                                                               & $\checkmark$                                                                                & $\checkmark$                                          & $\checkmark$                                     & $\cdot$       \\ 
\Xhline{1.5pt}
\end{tabular}}
\end{table}

Compared to W.E., SecDTD enables pre-Softmax benefits and avoids reliance on the complex Bitonic Sort, significantly enhancing its practicality for real-world inference.
Relative to CipherPrune, SecDTD not only provides pre-Softmax gains but also eliminates the need for expensive fine-tuning required for its thresholding method (though this does not impact inference time), thereby avoiding issues such as imprecise token dropping and, more critically, potential leakage risks. These advantages ensure both secure and efficient deployment in practice.

\begin{table}[h]
\centering
\caption{SecDTD Improvements through Fine-tuning.}\vspace*{-0.1in}
\label{Finetune}
\resizebox{82mm}{!}{
\begin{tabular}{cr|ccc}
\Xhline{1.5pt}
\multicolumn{2}{c|}{\textbf{Dataset}} 
& \textbf{CoLA}  & \textbf{STS-B} & \textbf{RTE}   \\ \hline

\multicolumn{2}{c|}{\textbf{Seq. Length}}                    
& 64 & 128 & 256 \\ \hline

\multicolumn{1}{c|}{\multirow{3}{*}{\rotatebox{90}{\textbf{LAN}}}}
& Baseline 
& 1370.46s & 3225.60s & 8464.14s \\

\multicolumn{1}{c|}{} 
& ours     
& 518.37s$\to$479.08s 
& 916.02s$\to$824.53s 
& 2355.26s$\to$2046.07s \\ \cline{2-5}

\multicolumn{1}{c|}{} 
& \multicolumn{1}{c|}{\textbf{Speedup}} 
& $\bm{2.54\times\to2.86\times}$ 
& $\bm{3.52\times\to3.91\times}$ 
& $\bm{3.59\times\to4.14\times}$ \\ \hline

\multicolumn{1}{c|}{\multirow{3}{*}{\rotatebox{90}{\textbf{WAN}}}}
& Baseline 
& 2377.48s & 5368.23s & 13475.92s \\

\multicolumn{1}{c|}{} 
& ours     
& 887.52s$\to$808.50s 
& 1533.24s$\to$1416.31s 
& 3837.30s$\to$3338.84s \\ \cline{2-5}

\multicolumn{1}{c|}{} 
& \multicolumn{1}{c|}{\textbf{Speedup}} 
& $\bm{2.68\times\to2.94\times}$ 
& $\bm{3.50\times\to3.79\times}$ 
& $\bm{3.51\times\to4.04\times}$ \\ \hline

\multicolumn{1}{c|}{\multirow{3}{*}{\rotatebox{90}{\textbf{MOB}}}}
& Baseline 
& 3393.21s & 7564.60s & 18638.96s \\

\multicolumn{1}{c|}{} 
& ours     
& 1250.05s$\to$1130.86s 
& 2149.83s$\to$2005.16s 
& 5346.08s$\to$4650.85s \\ \cline{2-5}

\multicolumn{1}{c|}{} 
& \multicolumn{1}{c|}{\textbf{Speedup}} 
& $\bm{2.71\times\to3.00\times}$ 
& $\bm{3.52\times\to3.77\times}$ 
& $\bm{3.49\times\to4.01\times}$ \\ 
\Xhline{1.5pt}
\end{tabular}}
\vspace*{-0.2in}
\end{table}

\subsubsection{Further Improvement with Fine-tune}
\label{Finetune sec}
Even without additional fine-tuning, SecDTD effectively reduces inference costs while preserving model accuracy through its dynamic and precise token drop mechanism. However, fine-tuning can further enhance performance by mitigating any potential accuracy loss, allowing for more aggressive token dropping while still maintaining baseline accuracy. As a result, both inference cost reduction and speedup are further improved with the incorporation of fine-tuning. As shown in Table~\ref{Finetune}, on datasets such as \textbf{CoLA}, \textbf{STS-B}, and \textbf{RTE}, which feature varying input sequence lengths, a simple fine-tuning process over \textbf{five epochs} leads to even greater speedup gains.

\subsection{SecDTD on BumbleBee Framework}
\label{SecDTD_BumbleBee}

As a general-purpose dynamic token drop method for secure Transformer inference, we replicate all experiments under the BumbleBee framework to further evaluate the effectiveness and generalizability of SecDTD, as shown in Table~\ref{BumbleBee_exp}. In this setting, SecDTD achieves similarly substantial cost reductions, with end-to-end inference speedup reaching up to $4.19\times$ compared to the baseline.
Consistent with its performance under the BOLT framework, SecDTD improves token drop effectiveness in the early stages of Transformer inference, particularly in components such as $\mathcal{S}$(Softmax) and $\mathcal{S}$($\times$V). The BumbleBee framework employs an NTT-accelerated polynomial multiplication protocol for matrix multiplication, resulting in a lower baseline cost for linear operations and, consequently, less room for optimization. This leads to a slightly reduced relative speedup compared to the BOLT framework. Nonetheless, SecDTD consistently delivers strong performance improvements across all cases, largely due to its efficient optimization of non-linear components, especially Softmax.

\begin{table*}[t]
\centering
\vspace*{-0.1in}\caption{SecDTD Performance under BumbleBee Framework without Fine-tuning.}\vspace*{-0.1in}
\label{BumbleBee_exp}
\begin{tabular}{ccr|cccccccc}
\Xhline{1.5pt}
\multicolumn{3}{c|}{\textbf{Dataset}}                                                                                                                                          & \textbf{CoLA}  & \textbf{SST-2} & \textbf{MRPC}  & \textbf{STS-B} & \textbf{QNLI}  & \textbf{QQP}   & \textbf{MNLI}  & \textbf{RTE}   \\ \hline
\multicolumn{1}{c|}{\multirow{3}{*}{\textbf{LAN}}}    & \multicolumn{1}{c|}{\multirow{2}{*}{\textbf{\begin{tabular}[c]{@{}c@{}}Inference\\ Time (s)\end{tabular}}}} & Baseline & 96.79          & 209.33         & 209.33         & 209.33         & 209.33         & 209.33         & 490.86         & 490.86         \\
\multicolumn{1}{c|}{}                                 & \multicolumn{1}{c|}{}                                                                                       & ours     & 45.28          & 65.24          & 62.47          & 66.02          & 90.38          & 65.92          & 117.04         & 145.21         \\ \cline{2-11} 
\multicolumn{1}{c|}{} &
\multicolumn{2}{c|}{\textbf{Speedup}} &
$\bm{2.14\times}$ &
$\bm{3.21\times}$ &
$\bm{3.35\times}$ &
$\bm{3.17\times}$ &
$\bm{2.32\times}$ &
$\bm{3.18\times}$ &
$\bm{4.19\times}$ &
$\bm{3.38\times}$ \\ \hline
\multicolumn{1}{c|}{\multirow{3}{*}{\textbf{WAN}}}    & \multicolumn{1}{c|}{\multirow{2}{*}{\textbf{\begin{tabular}[c]{@{}c@{}}Inference\\ Time (s)\end{tabular}}}} & Baseline & 645.85         & 1261.57        & 1261.57        & 1261.57        & 1261.57        & 1261.57        & 2957.89        & 2957.89        \\
\multicolumn{1}{c|}{}                                 & \multicolumn{1}{c|}{}                                                                                       & ours     & 347.24         & 466.35         & 444.88         & 462.70         & 610.95         & 481.42         & 751.82         & 906.58         \\ \cline{2-11} 
\multicolumn{1}{c|}{} &
\multicolumn{2}{c|}{\textbf{Speedup}} &
$\bm{1.86\times}$ &
$\bm{2.71\times}$ &
$\bm{2.84\times}$ &
$\bm{2.73\times}$ &
$\bm{2.06\times}$ &
$\bm{2.62\times}$ &
$\bm{3.93\times}$ &
$\bm{3.26\times}$ \\ \hline
\multicolumn{1}{c|}{\multirow{3}{*}{\textbf{MOBILE}}} & \multicolumn{1}{c|}{\multirow{2}{*}{\textbf{\begin{tabular}[c]{@{}c@{}}Inference\\ Time (s)\end{tabular}}}} & Baseline & 1150.30        & 2328.31        & 2328.31        & 2328.31        & 2328.31        & 2328.31        & 5591.46        & 5591.46        \\
\multicolumn{1}{c|}{}                                 & \multicolumn{1}{c|}{}                                                                                       & ours     & 588.67         & 807.89         & 772.39         & 807.59         & 1080.94        & 826.98         & 1351.27        & 1647.08        \\ \cline{2-11} 
\multicolumn{1}{c|}{} &
\multicolumn{2}{c|}{\textbf{Speedup}} &
$\bm{1.95\times}$ &
$\bm{2.88\times}$ &
$\bm{3.01\times}$ &
$\bm{2.88\times}$ &
$\bm{2.15\times}$ &
$\bm{2.82\times}$ &
$\bm{4.14\times}$ &
$\bm{3.39\times}$ \\ \Xhline{1.5pt}
\end{tabular}
\vspace*{-0.1in}
\end{table*}

\section{Discussion and Conclusion}
\label{conclusion}

\noindent\textbf{Why use half-and-half token drop in this work:}
First, both the length of HE ciphertexts (i.e., the number of slots) and operations or optimizations related to packed HE generally follow powers-of-two patterns, such as $\log_2$ or $2^n$. In practical implementations, dropping tokens to more than 50\% has no real effect, and dropping to 30\% (or any ratio between 50\% and 25\%) is effectively treated as a 50\% drop, with the remaining slots padded to satisfy ciphertext length constraints.
Therefore, employing a half-and-half token drop aligns optimally with this structure and improves computational efficiency.
Additionally, smaller ratios such as 12.5\% can be achieved through iterative application of token dropping (e.g., three times), offering greater flexibility.
Second, the half-and-half token drop also complements our design, which identifies the median and uses the average as a pivot, thereby improving the overall efficiency across different components of the system.

\vspace*{0.05in}
\noindent\textbf{Why not simply use the average as the median, instead of using it as a pivot to find the median:}
In brief, this simplification can lead to a 3\%–5\% drop in inference accuracy.
Although our MCN architecture shapes the score vector into a smooth, gradually decreasing trend—bringing the average closer to the median—this approximation is not exact. Using the average directly as the threshold for token dropping can introduce errors. If token selection continues to rely on this average, the discrepancy between the average and the true median may result in suboptimal decisions. The accumulation of such biased token drops can ultimately degrade model performance.

\begin{figure}[h]
\centering
\includegraphics[trim={0cm 0cm 0cm 0cm}, clip, scale=0.75]{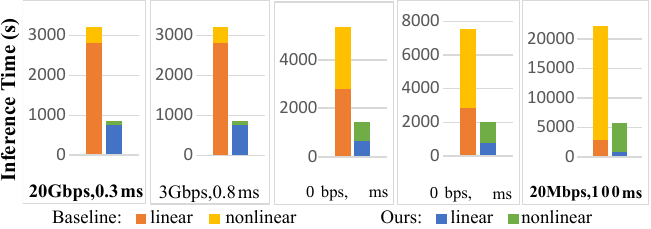}
\vspace*{-0.2in}
\caption{SecDTD Performance Breakdown from Extremely Fast to Extremely Slow Networks.} 
\label{Extreme}
\vspace*{-0.1in}
\end{figure}

\vspace*{0.05in}
\noindent\textbf{Performance of SecDTD in extreme network conditions:}
We further evaluated the end-to-end runtime of SecDTD under both extremely fast and extremely slow network conditions: one with 20 Gbps bandwidth and 0.3 ms latency (typical of high-speed dedicated networks), and the other with 20 Mbps bandwidth and 100 ms latency (commonly seen in IoT applications). This evaluation highlights the effectiveness and generalizability of our approach.
As shown in Figure~\ref{Extreme}, using the MRPC dataset within the BOLT framework as an example, the proportion of non-linear part in inference runtime increases steadily—and eventually dominates—as network conditions degrade from fast to slow. Nonetheless, SecDTD consistently achieves strong performance gains, primarily due to its efficient handling of both linear and non-linear components.
This aligns with our design goal: to provide a general-purpose dynamic token drop method for secure Transformer inference, without dependence on any specific deployment scenario for optimal effectiveness.

\vspace*{0.05in}\noindent\textbf{Generalizability of SecDTD:}
As a general-purpose dynamic token drop method for secure Transformer inference built on standard cryptographic primitives such as MPC and HE, SecDTD is theoretically applicable to all existing secure Transformer frameworks without requiring additional auxiliary components.
Currently, the most widely used secure Transformer frameworks include BOLT~\cite{pang2023bolt}, BumbleBee~\cite{lu2023bumblebee}, Iron~\cite{hao2022iron}, and NEXUS~\cite{zhang2024secure}. We have evaluated the performance of SecDTD within BOLT and BumbleBee. Although SecDTD can also be directly applied to Iron, we omit detailed evaluation. 
Nevertheless, theoretical analysis suggests that SecDTD would yield similar runtime reductions to those observed with BumbleBee.
For NEXUS, a ``non-interactive" framework, SecDTD could be applied if intermediate interaction is permitted, as our token drop mechanism relies on MPC. In this case, theoretical analysis also indicates that performance improvement comparable to those in BOLT could be achieved.
Our current implementation targets Transformer \textbf{encoders}, with BERT used as a representative model to demonstrate the effectiveness of our design. Transformer \textbf{decoders}, such as GPT~\cite{radford2018improving,radford2019language} and LLaMA~\cite{touvron2023llamaopenefficientfoundation}, share some similar architectural foundations with encoder-based models. Therefore, in principle, token drop techniques are also applicable in decoder settings. 
Overall, as a general-purpose dynamic token drop method for secure Transformer inference, SecDTD can be extended to larger Transformer models and larger datasets. Based on the above discussion, we anticipate that the performance gains would be even greater than those observed with BERT-base on GLUE. Further exploration of this potential by the community remains an interesting avenue for future work.

\vspace*{0.1in}
\noindent\textbf{CONCLUSION:}
In this work, we present SecDTD, a dynamic token drop framework designed to enable efficient and privacy-preserving Transformer inference. SecDTD enhances the effectiveness of token drop in secure settings by shifting the drop operation to earlier stages of inference. To achieve this, we introduce two key techniques: Max-Centric Normalization (MCN), a pre-Softmax scoring mechanism that evaluates token importance earlier and more efficiently; and OMSel, an secure oblivious median selection protocol that achieves up to 16.9$\times$ speedup over prior sorting-based methods while preserving security, obliviousness, and randomness. Extensive experiments across eight GLUE datasets and various network conditions demonstrate that SecDTD achieves up to 4.5$\times$ end-to-end inference acceleration without compromising model accuracy, providing a practical and scalable solution for secure Transformer inference.


\bibliographystyle{IEEEtran}
\bibliography{sample-base}

@misc{touvron2023llamaopenefficientfoundation,
      title={LLaMA: Open and Efficient Foundation Language Models}, 
      author={Hugo Touvron and Thibaut Lavril and Gautier Izacard and Xavier Martinet and Marie-Anne Lachaux and Timothée Lacroix and Baptiste Rozière and Naman Goyal and Eric Hambro and Faisal Azhar and Aurelien Rodriguez and Armand Joulin and Edouard Grave and Guillaume Lample},
      year={2023},
      eprint={2302.13971},
      archivePrefix={arXiv},
      primaryClass={cs.CL},
      url={https://arxiv.org/abs/2302.13971}, 
}

@article{zhang2025cipherprune,
  title={Cipherprune: Efficient and scalable private transformer inference},
  author={Zhang, Yancheng and Xue, Jiaqi and Zheng, Mengxin and Xie, Mimi and Zhang, Mingzhe and Jiang, Lei and Lou, Qian},
  journal={arXiv preprint arXiv:2502.16782},
  year={2025}
}

@inproceedings{asharov2013more,
  title={More efficient oblivious transfer and extensions for faster secure computation},
  author={Asharov, Gilad and Lindell, Yehuda and Schneider, Thomas and Zohner, Michael},
  booktitle={Proceedings of the 2013 ACM SIGSAC conference on Computer \& communications security},
  pages={535--548},
  year={2013}
}

@article{hendrycks2016gaussian,
  title={Gaussian error linear units (gelus)},
  author={Hendrycks, Dan and Gimpel, Kevin},
  journal={arXiv preprint arXiv:1606.08415},
  year={2016}
}

@inproceedings{wang2018glue,
  title={GLUE: A Multi-Task Benchmark and Analysis Platform for Natural Language Understanding},
  author={Wang, Alex and Singh, Amanpreet and Michael, Julian and Hill, Felix and Levy, Omer and Bowman, Samuel R},
  booktitle={Proceedings of the 2018 EMNLP Workshop BlackboxNLP: Analyzing and Interpreting Neural Networks for NLP},
  pages={353--355},
  year={2018},
  url={https://gluebenchmark.com/},
}

@inproceedings {spu,
    author = {Junming Ma and Yancheng Zheng and Jun Feng and Derun Zhao and Haoqi Wu and Wenjing Fang and Jin Tan and Chaofan Yu and Benyu Zhang and Lei Wang},
    title = {{SecretFlow-SPU}: A Performant and {User-Friendly} Framework for {Privacy-Preserving} Machine Learning},
    booktitle = {2023 USENIX Annual Technical Conference (USENIX ATC 23)},
    year = {2023},
    isbn = {978-1-939133-35-9},
    address = {Boston, MA},
    pages = {17--33},
    url = {https://www.usenix.org/conference/atc23/presentation/ma},
    publisher = {USENIX Association},
    month = jul,
}

@inproceedings{snoek2012practical,
  title={Practical Bayesian optimization of machine learning algorithms},
  author={Snoek, Jasper and Larochelle, Hugo and Adams, Ryan P},
  booktitle={NeurIPS},
  year={2012}
}

@inproceedings{tramer2016stealing,
  title={Stealing Machine Learning Models via Prediction APIs},
  author={Tramèr, Florian and et al.},
  year      = {2016},
  booktitle={USENIX Security 2016}
}

@inproceedings{shokri2017membership,
  title={Membership inference attacks against machine learning models},
  author={Shokri, Reza and et al.},
  year      = {2017},
  booktitle={IEEE S\&P 2017}
}

@inproceedings{nishide2007cmp,
  author    = {Takashi Nishide and Koki Ohta},
  title     = {Multiparty Computation for Interval, Comparison, and Equality without Bit-Decomposition Protocol},
  booktitle = {Public Key Cryptography - PKC 2007},
  year      = {2007},
  pages     = {343--360}
}

@inproceedings{catrina2010secure,
  author    = {Octavian Catrina and Amitabh Saxena},
  title     = {Secure Computation with Fixed-Point Numbers},
  booktitle = {Financial Cryptography and Data Security},
  year      = {2010},
  pages     = {35--50}
}

@inproceedings{demmler2015aby,
  author    = {Daniel Demmler and Thomas Schneider and Michael Zohner},
  title     = {ABY - A Framework for Efficient Mixed-Protocol Secure Two-Party Computation},
  booktitle = {NDSS},
  year      = {2015}
}

@inproceedings{bogdanov2014practical,
  title={A practical analysis of oblivious sorting algorithms for secure multi-party computation},
  author={Bogdanov, Dan and Laur, Sven and Talviste, Riivo},
  booktitle={Nordic Conference on Secure IT Systems},
  pages={59--74},
  year={2014},
  organization={Springer}
}

@inproceedings{beame2015finding,
  title={Finding the median (obliviously) with bounded space},
  author={Beame, Paul and Liew, Vincent and Pǎtra{\c{s}}cu, Mihai},
  booktitle={Automata, Languages, and Programming: 42nd International Colloquium, ICALP 2015, Kyoto, Japan, July 6-10, 2015, Proceedings, Part I 42},
  pages={103--115},
  year={2015},
  organization={Springer}
}

@article{dwork2014algorithmic,
  title={The algorithmic foundations of differential privacy},
  author={Dwork, Cynthia and Roth, Aaron and others},
  journal={Foundations and Trends{\textregistered} in Theoretical Computer Science},
  volume={9},
  number={3--4},
  pages={211--407},
  year={2014},
  publisher={Now Publishers, Inc.}
}

@inproceedings{rathee2021sirnn,
  title={Sirnn: A math library for secure rnn inference},
  author={Rathee, Deevashwer and Rathee, Mayank and Goli, Rahul Kranti Kiran and Gupta, Divya and Sharma, Rahul and Chandran, Nishanth and Rastogi, Aseem},
  booktitle={2021 IEEE Symposium on Security and Privacy (SP)},
  pages={1003--1020},
  year={2021},
  organization={IEEE}
}

@inbook{10.1145/3335741.3335756,
author = {Ben-Or, Michael and Goldwasser, Shafi and Wigderson, Avi},
title = {Completeness theorems for non-cryptographic fault-tolerant distributed computation},
year = {2019},
isbn = {9781450372664},
publisher = {Association for Computing Machinery},
address = {New York, NY, USA},
url = {https://doi.org/10.1145/3335741.3335756},
booktitle = {Providing Sound Foundations for Cryptography: On the Work of Shafi Goldwasser and Silvio Micali},
pages = {351–371},
numpages = {21}
}

@incollection{goldreich2019play,
  title={How to play any mental game, or a completeness theorem for protocols with honest majority},
  author={Goldreich, Oded and Micali, Silvio and Wigderson, Avi},
  booktitle={Providing Sound Foundations for Cryptography: On the Work of Shafi Goldwasser and Silvio Micali},
  pages={307--328},
  year={2019}
}

@inproceedings{karpukhin2020dense,
  title={Dense Passage Retrieval for Open-Domain Question Answering.},
  author={Karpukhin, Vladimir and Oguz, Barlas and Min, Sewon and Lewis, Patrick SH and Wu, Ledell and Edunov, Sergey and Chen, Danqi and Yih, Wen-tau},
  booktitle={EMNLP (1)},
  pages={6769--6781},
  year={2020}
}

@article{chen2021evaluating,
  title={Evaluating large language models trained on code},
  author={Chen, Mark and Tworek, Jerry and Jun, Heewoo and Yuan, Qiming and Pinto, Henrique Ponde De Oliveira and Kaplan, Jared and Edwards, Harri and Burda, Yuri and Joseph, Nicholas and Brockman, Greg and others},
  journal={arXiv preprint arXiv:2107.03374},
  year={2021}
}

@article{ott2018scaling,
  title={Scaling neural machine translation},
  author={Ott, Myle and Edunov, Sergey and Grangier, David and Auli, Michael},
  journal={arXiv preprint arXiv:1806.00187},
  year={2018}
}

@article{radford2018improving,
  title={Improving language understanding by generative pre-training},
  author={Radford, Alec and Narasimhan, Karthik and Salimans, Tim and Sutskever, Ilya and others},
  year={2018},
  publisher={San Francisco, CA, USA}
}

@article{radford2019language,
  title={Language models are unsupervised multitask learners},
  author={Radford, Alec and Wu, Jeffrey and Child, Rewon and Luan, David and Amodei, Dario and Sutskever, Ilya and others},
  journal={OpenAI blog},
  volume={1},
  number={8},
  pages={9},
  year={2019}
}

@article{vaswani2017attention,
  title={Attention is all you need},
  author={Vaswani, Ashish and Shazeer, Noam and Parmar, Niki and Uszkoreit, Jakob and Jones, Llion and Gomez, Aidan N and Kaiser, {\L}ukasz and Polosukhin, Illia},
  journal={Advances in neural information processing systems},
  volume={30},
  year={2017}
}

@inproceedings{devlin2019bert,
  title={Bert: Pre-training of deep bidirectional transformers for language understanding},
  author={Devlin, Jacob and Chang, Ming-Wei and Lee, Kenton and Toutanova, Kristina},
  booktitle={Proceedings of the 2019 conference of the North American chapter of the association for computational linguistics: human language technologies, volume 1 (long and short papers)},
  pages={4171--4186},
  year={2019}
}

@inproceedings{goyal2020power,
  title={Power-bert: Accelerating bert inference via progressive word-vector elimination},
  author={Goyal, Saurabh and Choudhury, Anamitra Roy and Raje, Saurabh and Chakaravarthy, Venkatesan and Sabharwal, Yogish and Verma, Ashish},
  booktitle={International Conference on Machine Learning},
  pages={3690--3699},
  year={2020},
  organization={PMLR}
}

@article{lu2023bumblebee,
  title={Bumblebee: Secure two-party inference framework for large transformers},
  author={Lu, Wen-jie and Huang, Zhicong and Gu, Zhen and Li, Jingyu and Liu, Jian and Hong, Cheng and Ren, Kui and Wei, Tao and Chen, WenGuang},
  journal={Cryptology ePrint Archive},
  year={2023}
}

@article{zhang2024secure,
  title={Secure transformer inference made non-interactive},
  author={Zhang, Jiawen and Yang, Xinpeng and He, Lipeng and Chen, Kejia and Lu, Wen-jie and Wang, Yinghao and Hou, Xiaoyang and Liu, Jian and Ren, Kui and Yang, Xiaohu},
  journal={Cryptology ePrint Archive},
  year={2024}
}

@article{pang2023bolt,
  title={BOLT: Privacy-Preserving, Accurate and Efficient Inference for Transformers},
  author={Pang, Qi and Zhu, Jinhao and M{\"o}llering, Helen and Zheng, Wenting and Schneider, Thomas},
  journal={Cryptology ePrint Archive},
  year={2023}
}

@inproceedings{zhang2024individual,
  title={From Individual Computation to Allied Optimization: Remodeling Privacy-Preserving Neural Inference with Function Input Tuning},
  author={Zhang, Qiao and Xiang, Tao and Xin, Chunsheng and Wu, Hongyi},
  booktitle={2024 IEEE Symposium on Security and Privacy (SP)},
  pages={101--101},
  year={2024},
  organization={IEEE Computer Society}
}

@article{hao2022iron,
  title={Iron: Private inference on transformers},
  author={Hao, Meng and Li, Hongwei and Chen, Hanxiao and Xing, Pengzhi and Xu, Guowen and Zhang, Tianwei},
  journal={Advances in Neural Information Processing Systems},
  volume={35},
  pages={15718--15731},
  year={2022}
}

@inproceedings {279898,
author = {Zhicong Huang and Wen-jie Lu and Cheng Hong and Jiansheng Ding},
title = {Cheetah: Lean and Fast Secure {Two-Party} Deep Neural Network Inference},
booktitle = {31st USENIX Security Symposium (USENIX Security 22)},
year = {2022},
isbn = {978-1-939133-31-1},
address = {Boston, MA},
pages = {809--826},
url = {https://www.usenix.org/conference/usenixsecurity22/presentation/huang-zhicong},
publisher = {USENIX Association},
month = aug
}

@inproceedings{cai2024mosaic,
  title={MOSAIC: A Prune-and-Assemble Approach for Efficient Model Pruning in Privacy-Preserving Deep Learning},
  author={Cai, Yifei and Zhang, Qiao and Ning, Rui and Xin, Chunsheng and Wu, Hongyi},
  booktitle={Proceedings of the 19th ACM Asia Conference on Computer and Communications Security},
  pages={1034--1048},
  year={2024}
}

@misc{sealcrypto,
    title = {{M}icrosoft {SEAL} (release 3.3)},
    howpublished = {\url{https://github.com/Microsoft/SEAL}},
    month = jun,
    year = 2019,
    note = {Microsoft Research, Redmond, WA.},
    key = {SEAL}
}

@misc{EzPC,
  author = "Microsoft Research",
  year =    "2019",
  title =  {Microsoft EzPC (Easy Secure Multi-party Computation)},
  url =    {https://www.microsoft.com/en-us/research/project/ezpc-easy-secure-multi-party-computation/},
}

@inproceedings{cai2022hunter,
  title={Hunter: He-friendly structured pruning for efficient privacy-preserving deep learning},
  author={Cai, Yifei and Zhang, Qiao and Ning, Rui and Xin, Chunsheng and Wu, Hongyi},
  booktitle={Proceedings of the 2022 ACM on Asia Conference on Computer and Communications Security},
  pages={931--945},
  year={2022}
}

@article{albrecht2021homomorphic,
  title={Homomorphic encryption standard},
  author={Albrecht, Martin and Chase, Melissa and Chen, Hao and Ding, Jintai and Goldwasser, Shafi and Gorbunov, Sergey and Halevi, Shai and Hoffstein, Jeffrey and Laine, Kim and Lauter, Kristin and others},
  journal={Protecting privacy through homomorphic encryption},
  pages={31--62},
  year={2021},
  publisher={Springer}
}

@inproceedings{brakerski2012fully,
  title={Fully homomorphic encryption without modulus switching from classical GapSVP},
  author={Brakerski, Zvika},
  booktitle={Annual Cryptology Conference},
  pages={868--886},
  year={2012},
  organization={Springer}
}

@inproceedings{sureshaby2,
  title={ABY2. 0: Improved mixed-protocol secure two-party computation},
  author={Patra, Arpita and Schneider, Thomas and Suresh, Ajith and Yalame, Hossein},
  booktitle={Proceedings of the USENIX Security},
  year={2021}
}

@inproceedings{mohassel2018aby3,
  title={ABY3: A mixed protocol framework for machine learning},
  author={Mohassel, Payman and Rindal, Peter},
  booktitle={Proceedings of the ACM SIGSAC},
  pages={35--52},
  year={2018}
}

@inproceedings{gentry2009fully,
  title={Fully homomorphic encryption using ideal lattices},
  author={Gentry, Craig},
  booktitle={Proceedings of the ACM STOC},
  pages={169--178},
  year={2009}
}

@article{brakerski2014efficient,
  title={Efficient fully homomorphic encryption from (standard) LWE},
  author={Brakerski, Zvika and Vaikuntanathan, Vinod},
  journal={SIAM Journal on computing},
  volume={43},
  number={2},
  pages={831--871},
  year={2014},
  publisher={SIAM}
}

@inproceedings{juvekar2018gazelle,
  title={$\{$GAZELLE$\}$: A low latency framework for secure neural network inference},
  author={Juvekar, Chiraag and Vaikuntanathan, Vinod and Chandrakasan, Anantha},
  booktitle={Proceedings of the USENIX Security},
  pages={1651--1669},
  year={2018}
}

@article{han2015learning,
  title={Learning both weights and connections for efficient neural networks},
  author={Han, Song and Pool, Jeff and Tran, John and Dally, William J},
  journal={arXiv preprint arXiv:1506.02626},
  year={2015}
}

@article{10.14778/3282495.3282499,
  title={Rafiki: Machine learning as an analytics service system},
  author={Wang, Wei and Wang, Sheng and Gao, Jinyang and Zhang, Meihui and Chen, Gang and Ng, Teck Khim and Ooi, Beng Chin},
  journal={arXiv preprint arXiv:1804.06087},
  year={2018}
}

@article{act1996health,
  title={Health insurance portability and accountability act of 1996},
  author={Act, Accountability},
  journal={Public law},
  volume={104},
  pages={191},
  year={1996}
}

@inproceedings{paillier1999public,
  title={Public-key cryptosystems based on composite degree residuosity classes},
  author={Paillier, Pascal},
  booktitle={Proceedings of the EUROCRYPT},
  pages={223--238},
  year={1999}
}

@article{fan2012somewhat,
  title={Somewhat practical fully homomorphic encryption.},
  author={Fan, Junfeng and Vercauteren, Frederik},
  journal={IACR Cryptol. ePrint Arch.},

  pages={144},
  year={2012},
  publisher={Citeseer}
}

@inproceedings{cheon2017homomorphic,
  title={Homomorphic encryption for arithmetic of approximate numbers},
  author={Cheon, Jung Hee and Kim, Andrey and Kim, Miran and Song, Yongsoo},
  booktitle={Proceedings of the ASIACRYPT},
  pages={409--437},
  year={2017}
}

@inproceedings{brassard1986all,
  title={All-or-nothing disclosure of secrets},
  author={Brassard, Gilles and Cr{\'e}peau, Claude and Robert, Jean-Marc},
  booktitle={Proceedings of the EUROCRYPT},
  pages={234--238},
  year={1986}
}

@article{shamir1979share,
  title={How to share a secret},
  author={Shamir, Adi},
  journal={Communications of the ACM},
  volume={22},
  number={11},
  pages={612--613},
  year={1979},
  publisher={ACm New York, NY, USA}
}

@inproceedings {244032,
author = {Pratyush Mishra and Ryan Lehmkuhl and Akshayaram Srinivasan and Wenting Zheng and Raluca Ada Popa},
title = {Delphi: A Cryptographic Inference Service for Neural Networks},
booktitle = {Proceedings of the USENIX Security},
year = {2020},
pages = {2505--2522}
}

@inproceedings{rathee2020cryptflow2,
  title={CrypTFlow2: Practical 2-party secure inference},
  author={Rathee, Deevashwer and Rathee, Mayank and Kumar, Nishant and Chandran, Nishanth and Gupta, Divya and Rastogi, Aseem and Sharma, Rahul},
  booktitle={Proceedings of the ACM SIGSAC},
  pages={325--342},
  year={2020}
}

@inproceedings{gilad2016cryptonets,
  title={Cryptonets: Applying neural networks to encrypted data with high throughput and accuracy},
  author={Gilad-Bachrach, Ran and Dowlin, Nathan and Laine, Kim and Lauter, Kristin and Naehrig, Michael and Wernsing, John},
  booktitle={Proceedings of the ICML},
  pages={201--210},
  year={2016}
}

@misc{speedtest,
  author =       "SPEEDTEST.net",
  year =         "2024",
  title =  {Market Analysis of Mobile and Fixed Broadband Speeds, US},
  url =    {https://www.speedtest.net/global-index/united-states},
  }

@misc{ChatGPT,
  author =       "openai.com",
  year =         "2025",
  title =  {ChatGPT: Get answers. Find inspiration. Be more productive.},
  url =    {https://openai.com/chatgpt/overview/},
  }

@misc{HIPAAcompliance2025,
  author =       "Timothy Shyu",
  year =         "2025",
  title =  {New legal developments herald big changes for HIPAA compliance in 2025.},
  url =    {https://www.reuters.com/legal/litigation/new-legal-developments-herald-big-changes-hipaa-compliance},
  }

@article{denecke2024transformer,
  title={Transformer models in healthcare: a survey and thematic analysis of potentials, shortcomings and risks},
  author={Denecke, Kerstin and May, Richard and Rivera-Romero, Octavio},
  journal={Journal of Medical Systems},
  volume={48},
  number={1},
  pages={23},
  year={2024},
  publisher={Springer}
}

@misc{cola,
  author =       "JeremiahZ",
  year={-},
  title =  {Model: bert-base-uncased-cola;mnli;qnli;rte;sst2;stsb},
  url =    {https://huggingface.co/JeremiahZ/bert-base-uncased-cola;mnli;qnli;rte;sst2;stsb},
  }

@misc{mrpc,
  author =       "Intel",
  year={-},
  title =  {Model: bert-base-uncased-mrpc},
  url =    {https://huggingface.co/Intel/bert-base-uncased-mrpc},
  }

@misc{qqp,
  author =       "gchhablani",
  year={-},
  title =  {Model: ert-base-cased-finetuned-qqp},
  url =    {https://huggingface.co/gchhablani/bert-base-cased-finetuned-qqp},
  }

@inproceedings{liu2017oblivious,
  title={Oblivious neural network predictions via minionn transformations},
  author={Liu, Jian and Juuti, Mika and Lu, Yao and Asokan, Nadarajah},
  booktitle={Proceedings of the ACM SIGSAC},
  pages={619--631},
  year={2017}
}

@inproceedings{brakerski2013packed,
  title={Packed ciphertexts in LWE-based homomorphic encryption},
  author={Brakerski, Zvika and Gentry, Craig and Halevi, Shai},
  booktitle={Proceedings of the PKC},
  pages={1--13},
  year={2013}
}

\appendices

\section{Analyzing Score Distributions for Efficient Median Selection}
\label{MCN_Distributions}

\begin{figure*}[t]
\centering
\includegraphics[trim={0cm 0cm 0cm 0cm}, clip, scale=0.6]{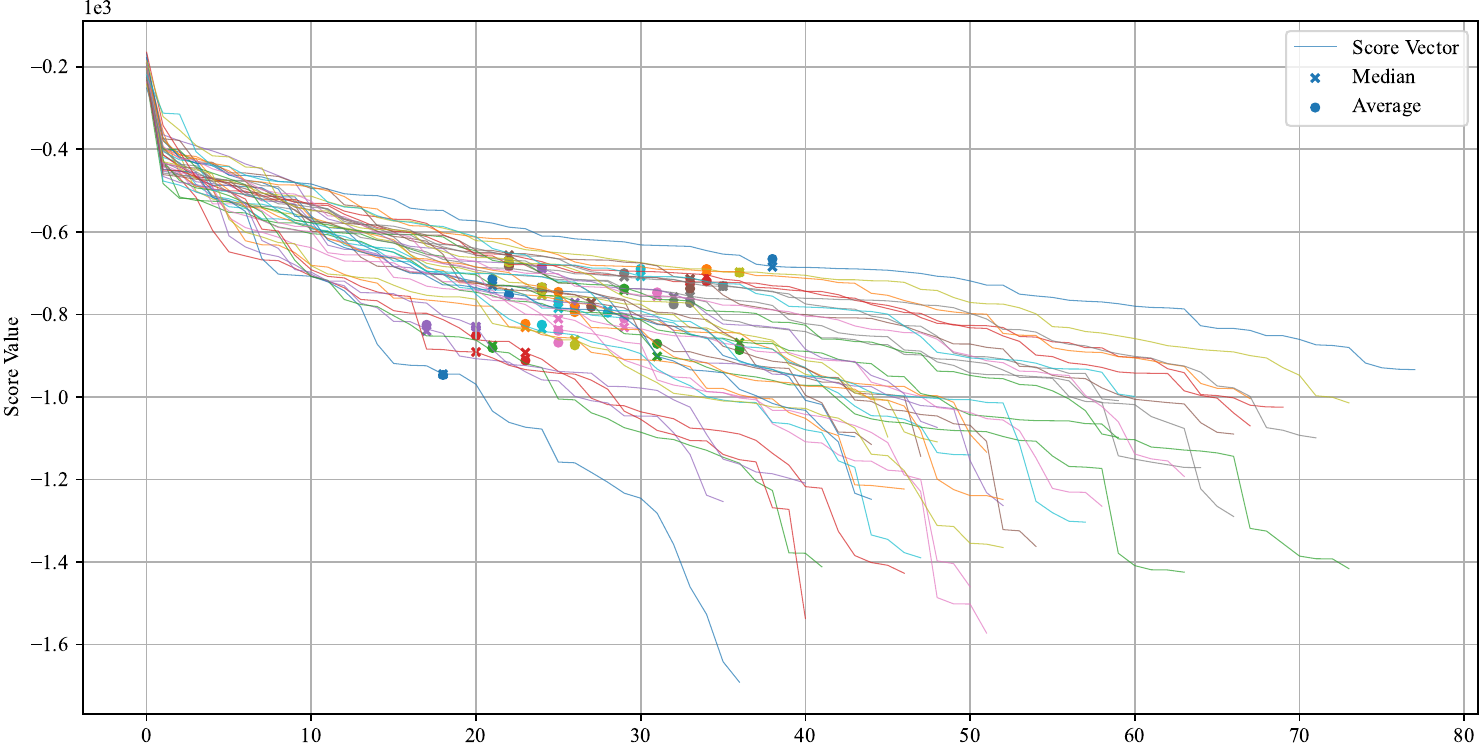}
\vspace*{-0.1in}
\caption{Mean and Median Distributions (40 Samples, MRPC Dataset).} 
\vspace*{-0.1in}
\label{MCN_avg_median}
\end{figure*}

The Figure~\ref{MCN_avg_median} illustrates the process of randomly selecting 40 score vector samples from the MRPC dataset during the Median Searching task, demonstrating the actual state of \textbf{OMSel} protocol \textbf{Step-1.2}. It is evident that the data processed through MCN exhibits a well-structured distribution, characterized by a smooth and gradually decreasing trend, resulting in a relatively uniform appearance. Notably, the middle segment of the curve remains particularly stable and evenly distributed. This characteristic ensures that the current \textbf{average} closely approximates the current \textbf{median}, facilitating rapid convergence when utilizing the average-as-pivot Median Selection. Consequently, this property enhances the efficiency and robustness of median searching.

\section{Complexity Analysis of OMSel}
\label{Complexity_Analysis}

\begin{figure}[h]
\centering
\includegraphics[trim={0cm 0cm 0cm 0cm}, clip, scale=0.56]{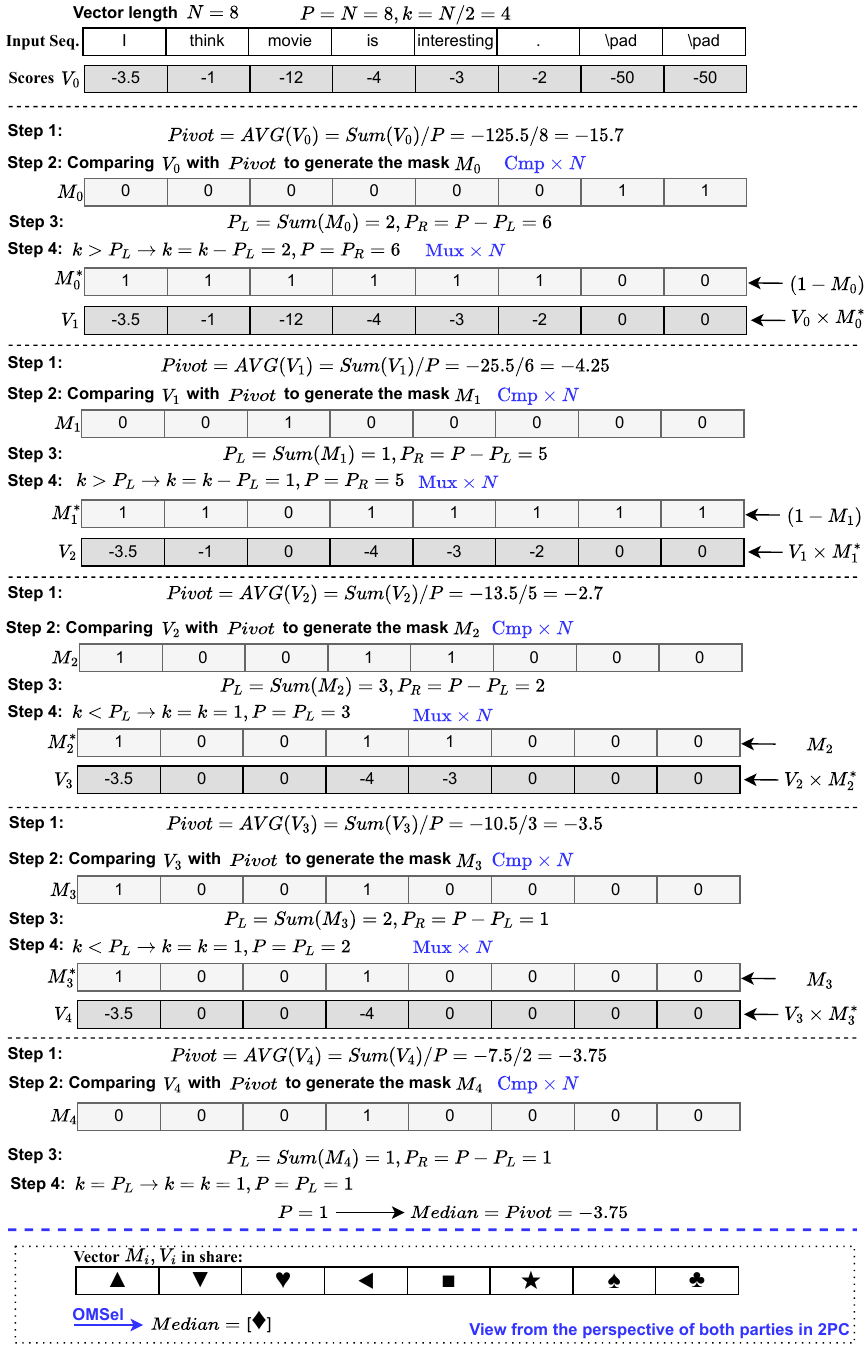}
\vspace*{-0.2in}
\caption{Toy example of OMSel (without RDM).} 
\vspace*{-0.1in}
\label{Toy_example_OMSel}
\end{figure}

As the toy example of OMSel shown in Figure~\ref{Toy_example_OMSel} illustrates, all $n$ elements are retained in each iteration to ensure obliviousness.
We then analyze the time complexity of OMSel, which consists of the initial random pivot step (RDM) followed by iterative refinements using the average as the pivot.

\vspace*{0.05in}
\noindent\textbf{Per-round Complexity.}
Each round requires:
\begin{itemize}[leftmargin=0in, itemindent=0.15in]
    \item \textbf{\#Cmp:} A pass to compare each element against the current pivot and generate corresponding masks: $\mathcal{O}(n)$;
    \item \textbf{\#Mux:} A pass to multiply all $n$ elements by the masks and update the score array: $\mathcal{O}(n)$;
\end{itemize}
Therefore, the complexity per round is:
\[
\boxed{C_r = \mathcal{O}(n)}
\]

\noindent\textbf{Number of Rounds.}
Let $R$ denote the number of refinement rounds required for the algorithm to converge. In the worst case—when each pivot is consistently the maximum or minimum value—$R$ can be as high as $\mathcal{O}(n)$. But, our approach mitigates this worst-case scenario by combining two mechanisms:
\begin{itemize}[leftmargin=0in, itemindent=0.15in]
    \item \textbf{Random Pivot Initialization:} The initial pivot $p_0$ is sampled uniformly at random from the input set. Thus, with high probability, $p_0$ is not an extreme value. This probabilistic guarantee helps separate most outliers (e.g., top or bottom 10\%) in the first round.
    \item \textbf{Average as Pivot:} In subsequent rounds, the the average of the elements is computed as the pivot. Under a well-structured distribution achieved by the proposed MCN scoring—characterized by a smooth and gradually decreasing trend—the average remains close to the center of mass, thereby staying near the true median and accelerating convergence.
\end{itemize}

By avoiding reliance on adversarial pivot choices and leveraging smoothing for stability, our approach prevents the degenerate behavior that leads to quadratic time complexity.

We analyze the number of rounds $R$ needed for the average-based pivot to converge to the median. Let $A = \{a_1, a_2, \dots, a_n\}$ be the input array of $n$ elements, and let $m$ denote its true median.
Since the average remains close to the true median, we assume that in each iteration, the pivot partitions the array such that the larger partition contains at most a fixed fraction $\alpha \in (0.5, 1)$ of the elements—for example, $\alpha = \frac{3}{5}$.
Thus, each round (logically) removes a constant fraction of the elements through proposed pseudo-partitioning. After $R$ rounds, the remaining number of elements is:
\[
n_r = n \cdot \alpha^R
\]
To achieve convergence (i.e., $n_r = 1$), the required number of rounds satisfies:
\[
n \cdot \alpha^R = 1 \quad \Rightarrow \quad R = \log_{1/\alpha}(n)
\]
Since $\alpha$ is a constant less than 1, the round complexity is:
\[
\boxed{R = \mathcal{O}(\log n)}
\]
\noindent\textbf{Total Complexity.}
Combining the per-round cost $\mathcal{O}(n)$ with $\mathcal{O}(\log n)$ rounds yields:
\[
\boxed{T(n) = \mathcal{O}(n \log n)}
\]

\end{document}